\documentclass[12pt]{article}

\usepackage{geometry}
\usepackage[onehalfspacing]{setspace}

\usepackage{amssymb,amsfonts,amsmath}
\usepackage{graphicx}
\usepackage{bbm}
\usepackage[T1]{fontenc}
\usepackage{microtype} 
\usepackage{xfrac}
\usepackage{enumitem}

\usepackage{booktabs}
\usepackage{tabularx}
\usepackage{threeparttable}
\usepackage{multirow}
\usepackage{longtable}
\usepackage[dvipsnames,table]{xcolor}
\usepackage{siunitx}
\usepackage{mathpazo}
\usepackage[scaled]{helvet}
\usepackage{courier}
\normalfont

\usepackage{float}
\usepackage{subcaption}
\usepackage{adjustbox}
\usepackage{pdflscape}  
\usepackage{soul}
\usepackage{ulem}
\newcounter{changenum}

\newcounter{rednum}

\usepackage{natbib}
\usepackage[pdftex,bookmarks,colorlinks]{hyperref}
\definecolor{darkpowderblue}{rgb}{0.0, 0.05, 0.5}
\hypersetup{
  colorlinks,
  citecolor=darkpowderblue,
  linkcolor=red,
  urlcolor=blue}
\usepackage{cleveref}

\definecolor{dpd}{rgb}{0.0, 0.05, 0.5}
\definecolor{editorange}{rgb}{0.85, 0.45, 0.0} 

\makeatletter
\newcommand\notsolarge{\@setfontsize\notsolarge{12}{14}}
\newcommand\footscript{\@setfontsize\footscript{8.5}{10}}
\makeatother

\renewenvironment{abstract}
 {\small
  \begin{center}
  \bfseries \abstractname\vspace{-.5em}\vspace{0pt}
  \end{center}
  \list{}{
    \setlength{\leftmargin}{1.6cm}
    \setlength{\rightmargin}{\leftmargin}}
  \item\relax}
 {\endlist}

\allowdisplaybreaks
\usepackage{xcolor}
\usepackage{colortbl}

\definecolor{ForestGreen}{RGB}{34,139,34}
\definecolor{dpd2}{rgb}{0.0, 0.035, 0.38}
\definecolor{RowAlt}{gray}{0.975}

\begin{document}
\sloppy


\title{\vspace*{1.2cm} \fontsize{22}{22}\selectfont
  \textbf{\color{dpd}
    \textls[-30]{Quantifying the Risk--Return Tradeoff in Forecasting}
  }
  \vspace*{1.65cm}}

\author{
  Philippe Goulet Coulombe\thanks{%
    D\'epartement des Sciences \'Economiques,
    \href{mailto:p.gouletcoulombe@gmail.com}{{\fontfamily{phv}\selectfont \color{dpd} p.gouletcoulombe@gmail.com}}. I thank Maximilian Göbel, Karin Klieber, and Anne Valder for helpful comments.
  }\\[-0.1cm]
  \textbf{ \vspace*{-0.25cm}
    {\fontfamily{phv}\selectfont \notsolarge Universit\'e du Qu\'ebec \`a Montr\'eal}
    \vspace*{1.2cm}}
}

\date{}
\maketitle


\begin{abstract}
\noindent Average forecast accuracy is not the same as forecast reliability. I treat forecast loss
differentials relative to a benchmark as a \textit{return} series. These return series are then evaluated using risk-adjusted performance measures from finance, including the Sharpe ratio, Sortino ratio, Omega ratio, and drawdown-based metrics.  I also introduce the \textit{Edge Ratio}
capturing a model’s propensity to deliver uniquely informative predictions relative
to the forecasting frontier. I apply this framework to U.S. macroeconomic forecasting,
comparing econometric benchmarks, machine learning models, a foundation model
(TabPFN), and the Survey of Professional Forecasters. While it is often feasible to beat professional forecasters in terms of average accuracy,
it is much harder to beat them on a \textit{risk-adjusted basis}. They rarely exhibit
catastrophic failures and often achieve high Edge Ratios, plausibly reflecting the value
of contextual judgment. Nonetheless, selected machine learning methods deliver
attractive risk profiles for specific targets. The framework naturally extends to
meta-analyses across targets, horizons, and samples, illustrated with a density forecast
evaluation and the M4 competition.
\end{abstract}

\thispagestyle{empty}

\clearpage
\setcounter{page}{1}


\section{Introduction}
\label{sec:intro}

Macroeconomic forecasters typically compare models using loss-based criteria such as root mean squared error (RMSE), mean absolute error (MAE), log scores, and equal predictive ability tests—most prominently the Diebold–Mariano test. These tools address two central questions: does a model reduce forecast errors relative to a benchmark, and is this improvement statistically significant? The literature on forecast evaluation is vast and sophisticated, but its unifying focus remains average performance gains and testing whether they differ from zero.

Central banks, finance ministries, and asset managers invest heavily in forecasting infrastructure: human capital, computing resources, and model development pipelines. The return on this investment is not a lower RMSE \textit{number} but a more reliable \textit{path of forecast losses}—consistently reduced uncertainty about the economic outlook. These users have naturally asymmetric preferences: they tolerate, even appreciate, episodes where a model dramatically outperforms expectations, but they cannot afford episodes of severe underperformance. A machine learning model that beats the benchmark by 15\% on average yet falters badly during the 2008 financial crisis or the 2022 inflation surge offers a less attractive risk-return profile than its average performance suggests.

It is not realistic to \emph{require} a new model to dominate a benchmark at every realization. Yet this unavoidable fact introduces risk. Two models may deliver similar RMSE ratios and both beat a benchmark with statistical significance while exhibiting markedly different patterns of gains and losses over time, resulting in different risk profiles. This distinction is difficult to discern from the results tables that typically accompany forecasting studies.

\vskip 0.4cm
\noindent \textbf{Forecasting Models as a Risky Investment.} This paper imports risk-adjusted performance measures from finance to address this gap. In portfolio management, returns are never judged by their mean alone. The \textit{Sharpe ratio} \citep{sharpe1966mutual} is the ratio of average return to return volatility, measuring improvement per unit of risk. The \textit{Sortino ratio} \citep{sortino1991downside} refines this by focusing on the semi-standard deviation---the volatility of negative returns only---which penalizes episodes where the model underperforms the benchmark while remaining agnostic to upside variability. The \textit{Omega ratio} \citep{keating2002universal} captures the full distribution of gains versus losses, comparing total upside to total downside without imposing distributional assumptions; this is particularly appealing in the forecasting context, where the distribution of loss differentials can be highly non-Gaussian. Finally, \textit{maximum drawdown} tracks the worst cumulative underperformance from peak to trough, identifying prolonged episodes of sustained forecast failure \cite{magdon2004analysis,chekhlov2005drawdown}.

These metrics have direct analogues for forecast evaluation. Defining the ``return'' at time $t$ as the loss differential $r_t = L_t^{\text{benchmark}} - L_t^{\text{model}}$, one can compute Forecast Sharpe, Forecast Sortino, and Forecast Omega ratios that summarize not just how much a model improves upon the benchmark, but how \textit{reliably} it does so. As I show in the results, some models delivering very similar MAE/RMSE improvements, and sometimes similar levels of statistical significance, feature strikingly different risk-adjusted ratios. That is, while there could be ambiguity with respect to standard metrics, the risk-adjusted ones provide a sharper answer as to which model is likely operationally preferable.

Importantly, the framework is not completely estranged from traditional tests of statistical significance. In fact, the simplest ratio---the Sharpe ratio---coincides with the Diebold-Mariano $t$-statistic (up to a scaling constant) when forecast errors are serially uncorrelated. Moreover, I show in Section~\ref{sec:dm_connection} that the Diebold-Mariano test can itself be interpreted as a utility-based statistic: the long-run variance in its denominator penalizes not only for the variance of loss differentials but also for their autocorrelation, implying that persistent deviations in performance are treated as more costly than transient failures. From an institutional standpoint, this asymmetric treatment of clustered outcomes has intuitive appeal: a prolonged streak of underperformance is more likely to materialize into sub-optimal policy decisions.

\vskip 0.4cm
\noindent \textbf{The Edge Ratio.} I also introduce the Edge Ratio, a novel metric that measures whether a model delivers unique predictive advantage relative to all available alternatives. Unlike benchmark-relative metrics, the Edge Ratio tracks how often a model reaches the forecasting frontier and how decisively it does so, while penalizing the magnitude of regret when it fails to be the best. Conceptually, it is an Omega ratio with a moving benchmark: at each point in time, the benchmark is the best-performing alternative model rather than a fixed reference. A model with high Edge Ratio provides value that cannot be replicated by other methods in the forecaster's toolkit without incurring a significant amount of realized regret with respect to competitors. 

\vskip 0.4cm
\noindent \textbf{Meta-Analyses.} Forecasting studies routinely evaluate new methods across many targets, horizons, geographies, and out-of-sample periods. Results are typically summarized by averages of, e.g., RMSEs, but substantial heterogeneity often lurks beneath. From a model development perspective, knowing which method wins on average is useful, but so is knowing which methods are robust---and which ones excel on a few targets while failing catastrophically on others. This matters especially because published studies tend to emphasize favorable cases; the question is whether these gains come at the cost of severe underperformance elsewhere. The same risk-adjusted metrics apply here: the Sortino ratio, for instance, asks how large the average improvement is relative to the magnitude of the worst failures across the design space. For practitioners moving from research to implementation, this is precisely the relevant question: if the setup differs slightly from the original study, how much variability in performance should one expect? Since model devolvement is itself a risky investment, such metrics can bring useful information to the table.

\vskip 0.4cm
\noindent \textbf{Three Applications.} I develop this framework and apply it to three settings.

\vskip 0.15cm
\noindent \textit{ML/AI versus SPF.} The first application evaluates risk-adjusted forecast performance over time. The exercise follows a standard recursive out-of-sample design from 2007 to the present, split into pre- and post-COVID eras. The model pool is inspired by \cite{dual} and includes standard econometric benchmarks (autoregressive, factor-augmented), regularized linear methods (Ridge, Kernel Ridge), and established ML approaches (Random Forests, gradient boosting, neural networks). Three recent contenders are of particular interest: (i) LGB+, a hybrid boosting algorithm that alternates tree ensembles with linear corrections \cite{lgbplus}; (ii) the Survey of Professional Forecasters (SPF), long known to be difficult to beat; and (iii) TabPFN, a foundation model trained exclusively on synthetic data to mitigate lookahead bias \cite{hollmann2022tabpfn}. This setup allows one to ask: on a risk-adjusted basis, what performs best---professional forecasters, traditional machine learning, or recent AI developments such as foundation models? The latter are admittedly more black-box in nature, but if they deliver strong returns with limited downside, practitioners may be willing to trade some interpretability for reliability. 

For GDP at $h=1$, several models---including neural networks and Kernel Ridge Regression---deliver larger average MSE reductions than the SPF. Yet the SPF's risk-adjusted performance is substantially stronger: its Sortino and Omega ratios lead most competitors by a wide margin, reflecting exceptionally contained downside risk. The foundation model TPFN also gets a fair shot. At longer horizons, the SPF dominates unambiguously once risk is accounted for. The Edge Ratio reveals that both the SPF and factor models sporadically deliver unique information at short horizons; crucially, when the SPF does not come out on top, its regret is contained. Tree-based models (RF, LGB, LGB+) tend to be more robust than neural networks or factor models---their returns are not as severely deflated by risk adjustment \citep{coulombe2022how}. For inflation, the Hemisphere Neural Network (HNN) stands out: it matches the SPF in raw returns but dominates on risk-adjusted metrics, and uniquely retains positive returns post-2021 when most ML models break down. This robustness to regime change is rare in the literature.

\vskip 0.15cm
\noindent \textit{Density forecasting meta-analysis.} Revisiting the forecasting exercise of \citet{GCFK}, HNN and BART emerge as the strongest point forecasters on a risk-adjusted basis, with BART achieving a marginally higher Sharpe ratio (0.97 vs.\ 0.85) thanks to lower performance volatility, while both share identical Sortino ratios (1.62). For density forecasts (log scores), HNN dominates decisively: its gains are larger, more stable, and less prone to catastrophic failures. The meta-analysis metrics confirm quantitatively what qualitative inspection suggests: some models are systematically more robust than others.

\vskip 0.15cm
\noindent \textit{M4 competition.} The M4 forecasting competition \citep{makridakis2018m4,makridakis2020m4} provides 48,000 monthly series evaluated under standardized protocols, using the Mean Absolute Scaled Error (MASE) and Overall Weighted Average (OWA) as primary metrics. Here, I find that average and risk-adjusted performance broadly align: models that rank highly by MASE also rank highly by risk-adjusted ratio. This concordance makes sense---methods that perform well across diverse series and data-generating processes are unlikely to fail catastrophically on any given target. Interestingly, the top-ranked models also exhibit substantial Edge Ratios, suggesting they are not merely conservative but occasionally deliver decisive gains unavailable to competitors.

\vskip 0.4cm
\noindent \textbf{Related Literature.} This paper connects three strands of research.

\vskip 0.15cm
\noindent \textit{Forecast evaluation and predictive ability tests.} The modern literature begins with the Diebold-Mariano test \citep{diebold1995comparing}, which evaluates whether mean loss differentials are statistically different from zero. Subsequent work refined this framework in multiple directions: \cite{west1996asymptotic} accounted for parameter estimation uncertainty; \cite{harvey1997testing} improved small-sample properties; \cite{giacomini2006tests} developed tests of conditional predictive ability. For multiple model comparisons, \cite{white2000reality} introduced the reality check for data snooping, refined by \cite{hansen2005test} and culminating in the Model Confidence Set \citep{hansen2011model}. A parallel literature emphasized that loss function choice matters for forecast rankings \citep{christoffersen1997optimal,granger1999outline,patton2007properties}. These procedures focus on statistical significance. The risk-adjusted metrics proposed here complement them by capturing the magnitude and stability of forecast gains---dimensions not always transparent when the focus is on whether a difference is significant. Such information is visible in cumulative sum of squared error plots, but can be difficult to summarize efficiently; these ratios provide useful summary statistics of this distribution.

\vskip 0.15cm
\noindent \textit{Scoring rules and decision-theoretic foundations.} The framework builds on proper scoring rules \citep{gneiting2007strictly,gneiting2011making}: the loss function generating the return series can be any consistent scoring function---squared error, absolute error, CRPS, or log score---each yielding different risk-adjusted rankings. The approach connects to regret-based decision theory \citep{bell1982regret,loomes1982regret,savage1951theory} and the broader argument that forecast evaluation should link to economic decisions \citep{granger2000economic,west2006forecast}. The Sortino ratio implicitly reflects regret-averse preferences. The Edge Ratio connects more directly to minimax regret \citep{savage1951theory}: by measuring performance relative to the best available model at each point in time, it quantifies how much a forecaster leaves on the table by not having access to the ex-post optimal choice---precisely the regret concept central to robust decision-making under uncertainty. Importantly, by evaluating the full path of loss differentials rather than conditioning on extreme outcomes, this avoids the ``forecaster's dilemma'' identified by \cite{lerch2017forecaster}.

\vskip 0.15cm
\noindent \textit{Machine learning in macroeconomic forecasting.} Factor models established the benchmark for high-dimensional macroeconomic prediction \citep{stock2002forecasting}. A large subsequent literature has shown that machine learning methods can improve forecast accuracy in various settings \cite{medeiros2021forecasting,coulombe2022how}. More recently, a new wave of models has pushed this agenda further, including foundation models for tabular time series such as TabPFN \citep{hollmann2022tabpfn} and applications of large language models to macroeconomic forecasting \citep{CarrieroPettenuzzoShekhar2024_MacroLLM,bybee2023surveying,alam2026chatmacro}. Yet nearly all of this literature evaluates performance using average loss measures such as RMSE or MAE and tests of statistical significance. As \cite{rossi2021forecasting} emphasizes, forecast performance can be highly unstable across time, and models that excel in one period may fail dramatically in another. The question posed in this paper is different: do these methods differ in their \textit{risk} profiles? In particular, are machine learning forecasts more volatile or more prone to catastrophic underperformance than traditional econometric models or professional forecasters, and which approaches offer the most attractive return--risk tradeoff?

\vskip 0.4cm
\noindent \textbf{Outline.} Section~\ref{sec:methodology} develops the risk-adjusted forecast evaluation framework. Section~\ref{sec:showdown} presents the main empirical application comparing ML models, a foundation model, and professional forecasters. Section~\ref{sec:meta} reports two additional meta-analyses. Section~\ref{sec:conclusion} concludes.


\section{Risk-Adjusted Forecast Evaluation}
\label{sec:methodology}

This section develops a risk-adjusted framework for forecast evaluation. I proceed in four steps. First, I define forecast gains relative to a benchmark and interpret them as a return series. Second, I introduce risk-adjusted performance metrics—Sharpe, Sortino, Omega, and maximum drawdown—borrowed from the evaluation of asset returns and trading strategies, and adapted here to the context of forecast losses. Third, I introduce the \emph{Edge Ratio}, a novel metric designed to elicit whether a model delivers a unique predictive advantage relative to the full forecasting frontier. Finally, I extend the same logic to a meta-analysis setting, aggregating performance across targets, horizons, and evaluation designs to assess the robustness of model improvements beyond a single forecasting problem.

\subsection{Forecast Gains and the Return Analogy}
\label{sec:forecast_gains}

Let $t = 1, \dots, T$ index the forecast evaluation periods.  For a model $M$ and benchmark $B$, let
\[
    L_t^{M} = L(y_t, \hat{y}_t^{M}), \qquad L_t^{B} = L(y_t, \hat{y}_t^{B}),
\]
where $L(\cdot, \cdot)$ is any loss function (squared error, absolute error, log score).  Define the {forecast gain} of $M$ over $B$ at time $t$ as
\[
    r_t = L_t^{B} - L_t^{M}.
\]
Here $r_t > 0$ means model $M$ beats the benchmark at time $t$, while $r_t < 0$ means model $M$ underperforms.  The sequence $\{r_t\}_{t=1}^{T}$ is the ``return'' series from switching from benchmark $B$ to model $M$. The average return
\[
    \bar{r} = \frac{1}{T} \sum_{t=1}^{T} r_t,
\]
which corresponds to the improvement in average loss.  However, $\bar{r}$ alone says nothing about the \textit{volatility} of gains, the \textit{asymmetry} between good and bad periods, or the \textit{worst cumulative underperformance} experienced along the way. Yet the institutions that rely on these forecasts—central banks, governments, and institutional investors—care precisely about these dimensions. What matters to them is not only whether a model improves performance on average, but how those improvements and deteriorations materialize over time: whether gains are stable or erratic, whether losses are concentrated or persistent, and how severe the worst episodes of underperformance are.


\subsection{Risk-Adjusted Forecast Metrics}
\label{sec:risk_metrics}

I begin with risk-adjusted performance metrics borrowed from asset pricing and trading-strategy evaluation, introduced in increasing order of sophistication.

\paragraph{Forecast Sharpe Ratio.}
The Sharpe ratio \citep{sharpe1966mutual, sharpe1994sharpe} is the canonical risk-adjusted performance measure in finance. I adapt it to forecast evaluation by defining the {Forecast Sharpe ratio} as
\[
    \text{Sharpe} = \frac{\bar{r}}{s_r},
    \qquad \text{where} \qquad
    s_r^2 = \frac{1}{T-1} \sum_{t=1}^{T} (r_t - \bar{r})^2.
\]
This measure captures the average reduction in loss per unit of volatility of that reduction. For example, a model with $\text{Sharpe}=0.5$ delivers half a unit of average gain per unit of standard deviation, indicating relatively stable improvements. By contrast, a model with $\text{Sharpe}=0.1$ may exhibit a positive average gain but with highly volatile performance over time.

\paragraph{Forecast Sortino Ratio.}
The Sortino ratio \citep{sortino1991downside, sortino1994performance} addresses a key limitation of the Sharpe ratio, which treats upside and downside volatility symmetrically. In contrast, forecast users often have asymmetric preferences: periods of outperformance are typically tolerated, whereas periods of underperformance are more consequential. The {Forecast Sortino ratio}, defined as
\[
    \text{Sortino} = \frac{\bar{r}}{s_{\text{down}}},
    \qquad \text{where} \qquad
    s_{\text{down}} = \sqrt{\frac{1}{T} \sum_{t=1}^{T} (r_t^{-})^2},
    \quad r_t^{-} = \min(r_t, 0),
\]
constructs the denominator by focusing on what truly keeps forecasters awake at night: the average magnitude of \textit{underperformance}. The measure therefore isolates episodes in which the model loses to the benchmark. A high Sortino ratio indicates that such losses are infrequent and/or limited in magnitude. In this sense, the Sortino ratio is better aligned with the preferences of forecasters and policy makers than symmetric volatility-based performance measures.

Other downside-focused performance measures are possible, including drawdown-based ratios such as the Calmar and Sterling ratios, which penalize sustained cumulative underperformance rather than pointwise losses. I focus on the Sortino ratio because it provides a simple, observation-level measure of asymmetric risk that maps directly to forecast loss differentials, without introducing path dependence. In this setting, it offers a natural and interpretable summary of downside forecast risk.

\paragraph{Forecast Omega Ratio.}
The Omega ratio \citep{keating2002universal} captures the entire return distribution without parametric assumptions:
\[
    \Omega = \frac{\frac{1}{T} \sum_{t=1}^{T} r_t^{+}}{\frac{1}{T} \sum_{t=1}^{T} |r_t^{-}|} = \frac{\text{Average Upside}}{\text{Average Downside}}.
\]
This captures the overall balance between good and bad forecast episodes.  $\Omega > 1$ indicates more upside than downside; $\Omega < 1$ indicates the reverse.  Unlike the Sharpe or Sortino ratios, the Omega ratio imposes minimal structure on the underlying distribution, as it does not normalize returns by a variance or semi-variance term. This is particularly appealing in this setting, where the objects of interest are differences in forecast losses, whose empirical distribution need not be symmetric, light-tailed, or even approximately Gaussian in small samples. Moreover, since the goal is not statistical inference on a scalar performance index but rather a comparative assessment of realized forecast paths, avoiding moment-based normalization can be viewed as a feature rather than a limitation. The cost is, however, to obtain a metric with slightly less interpretable units.

\paragraph{Maximum Drawdown.}
Maximum drawdown is a path-dependent risk measure capturing the largest cumulative loss from a historical peak \citep{magdon2004analysis, chekhlov2005drawdown}. Define cumulative gains as $R_0 = 0$ and $R_t = \sum_{s=1}^{t} r_s$ for $t \ge 1$. The running maximum is $M_t = \max_{0 \le u \le t} R_u$, and the drawdown at time $t$ is $\text{DD}_t = M_t - R_t$. Maximum drawdown is then defined as $\text{MaxDD} = \max_{1 \le t \le T} \text{DD}_t$.

Maximum drawdown summarizes the depth of the most severe cumulative underperformance episode along the forecast path. A model may therefore display strong average performance while still experiencing prolonged or economically meaningful periods of underperformance. Such episodes are particularly relevant in macroeconomic forecasting—for example, during the post-pandemic inflation surge of 2021–2022—where sustained forecast errors can have material policy implications. As such, maximum drawdown provides a complementary dimension of forecast evaluation that is not captured by average loss measures.

\paragraph{Remarks on Coherence and the Choice of Loss Function.}
{The risk-adjusted metrics developed here evaluate forecasts \emph{ex post}, after they have been produced under some loss function (typically squared error). A natural question arises: if a user's preferences are asymmetric---as the Sortino ratio implicitly assumes---should the forecast itself have been optimized under a different loss? \citet{gneiting2011making} establishes that coherent point forecast evaluation requires matching the scoring function to the functional being targeted. A scoring function is consistent for the mean if and only if it is a Bregman function, while consistency for a quantile requires the asymmetric piecewise linear loss, and consistency for an expectile requires asymmetric squared error \citep{newey1987asymmetric, ehm2016quantiles}.

This tension is endemic to forecast evaluation in practice, where forecasters rarely receive explicit directives and users rarely articulate their loss functions. The framework does not resolve this ambiguity---it instead takes the loss function as given and asks a complementary question: \emph{conditional on this choice}, how stable are the resulting gains? The Sortino ratio's asymmetric treatment of the comparison distribution does not conflict with the symmetry of squared error as a scoring rule: the loss function determines \emph{what} the forecasters are compared on, while the Sortino ratio characterizes \emph{how reliably} one dominates the other. In principle, a user with strong asymmetric preferences could define $L_t$ using quantile or expectile loss throughout, yielding a fully coherent evaluation; I focus on squared and absolute error because they remain dominant in macroeconomic forecasting practice.
}

\paragraph{Serial Dependence and Short Samples.}
The Sharpe, Sortino, and Omega ratios defined above use contemporaneous standard or semi-standard deviations rather than HAC-adjusted long-run variances. This is a deliberate choice: these ratios are intended as descriptive summaries of the realized gain distribution, not as test statistics requiring asymptotic pivotality. However, when loss differentials exhibit substantial autocorrelation---as is common with direct multi-step forecasts at $h \ge 2$, where overlapping forecast horizons mechanically induce serial dependence---the effective sample size is smaller than the nominal one, and the ratios become noisier estimators of the underlying risk--return profile. This is particularly relevant for the post-COVID evaluation windows in the present application, which span approximately 13--17 quarters. I therefore present these localized ratios as suggestive evidence of regime-specific performance rather than as precise estimates, and I report first-order autocorrelation coefficients $\rho(1)$ alongside each table to allow readers to gauge the degree of temporal dependence.


\subsection{Connection to Diebold-Mariano}
\label{sec:dm_connection}

The Diebold-Mariano (DM) statistic tests whether the mean loss differential $\bar{r}$ is significantly different from zero.  Under standard regularity conditions,
\[
    \text{DM} = \frac{\bar{r}}{\sqrt{\widehat{\text{LRV}}/T}},
\]
where $\widehat{\text{LRV}}$ is a HAC estimator of the long-run variance:
\[
    \text{LRV} = \gamma_0 + 2 \sum_{k=1}^{K} \gamma_k, \qquad \gamma_k = \text{Cov}(r_t, r_{t-k}).
\]
This shows that DM is formally a $t$-statistic where ``risk'' corresponds to the long-run variance.  If the loss differential has no temporal dependence ($\gamma_k = 0$ for $k > 0$), then
\[
    \text{DM} = \sqrt{T} \cdot \frac{\bar{r}}{s_r} = \sqrt{T} \cdot \text{Sharpe}.
\]
Thus, the Sharpe ratio is the normalized DM statistic when forecast errors are serially uncorrelated. Positive autocorrelation in $r_t$, which is typically what is observed in empirical macroeconomic studies, increases the long-run variance and penalizes the DM statistic.

\paragraph{DM as a Risk-Adjusted Metric.}
From a decision-theoretic standpoint, the DM statistic can itself be viewed as a risk-adjusted measure of forecasting performance. The denominator penalizes not only for the variance of loss differentials, but also for their serial correlation. When $\bar{r} > 0$, this additional penalty is economically meaningful: higher autocorrelation inflates the long-run variance and shrinks the DM statistic toward zero, reducing the statistical evidence for outperformance. Symmetrically, when $\bar{r} < 0$, the same mechanism makes a negative DM statistic less negative, effectively attenuating the measured severity of underperformance. In both cases, persistent loss differentials are treated as noisier evidence of a model's true relative performance. From a practical standpoint, if a model underperforms the benchmark and does so persistently over multiple periods, this constitutes a more severe failure than if the same cumulative underperformance were dispersed randomly across the evaluation window. Prolonged episodes of inferior forecasting—such as those observed during structural breaks or regime shifts—are arguably more costly from a practical perspective, whether for policy makers relying on the forecasts or for researchers assessing model credibility.

\paragraph{Practical Implications.}
This interpretation suggests that two models delivering identical reductions in average loss (e.g., RMSE) may nevertheless differ in their reliability as measured by the DM statistic. The model with more significant DM results exhibits loss differentials that are either less variable, less persistent, or both—implying more consistent outperformance across the evaluation sample. From this perspective, preferring the model with the higher DM $t$-statistic is not merely a matter of statistical significance, but reflects a preference for more dependable forecasting gains.

\paragraph{Limitations.}
The DM framework provides a principled approach to forecast comparison, but it was designed for hypothesis testing on average predictive ability, not for characterizing the risk profile of forecast gains. It treats upside and downside deviations symmetrically, since both contribute to the long-run variance used in the test statistic. When the goal shifts from ``is Model A significantly better on average?''\ to ``how reliably does Model A outperform, and how badly can it fail?'', the asymmetric risk measures developed above provide a natural complement.


\subsection{Edge Ratio:  Eliciting Unique Predictive Advantage}
\label{sec:edge_ratio}

Beyond benchmark-relative performance, I find it useful in practice to monitor whether a forecasting model delivers a \emph{unique} advantage relative to the full set of available alternatives. I therefore introduce one additional ratio—the \emph{Edge Ratio}—designed to track how often and how strongly a model reaches the forecasting frontier, while penalizing how far it falls behind the best competing model when it does not.

Conceptually, the Edge Ratio can be viewed as an Omega ratio with a \emph{moving benchmark}. Rather than comparing a model to a fixed reference, the benchmark is defined at each point in time as the ex post best-performing alternative among all competing models, excluding the model under evaluation. In this sense, the Edge Ratio evaluates performance against an explicitly adversarial benchmark: the model is rewarded only when it dominates all competitors, and penalized whenever another model performs better.

Let $L_{M,t}$ denote the loss of model $M$ at time $t$, and define the contemporaneous frontier loss as
\[
    L^{\star}_t = \min_{j \neq M} L_{j,t}.
\]
The \emph{edge} of model $M$ at time $t$ is then
\[
    e_{M,t} = L^{\star}_t - L_{M,t},
\]
so that $e_{M,t} > 0$ indicates that model $M$ attains the forecasting frontier at time $t$, while $e_{M,t} < 0$ measures the extent to which it underperforms relative to the best available alternative.

To mirror the upside–downside decomposition used in earlier performance measures, define \emph{edge wins} and \emph{edge regrets} as
\[
    e_{M,t}^{+} = \max(e_{M,t}, 0),
    \qquad
    e_{M,t}^{-} = \max(-e_{M,t}, 0).
\]
Edge wins capture how much the model outperforms the frontier when it is best, while edge regrets capture how costly it is, in hindsight, to have used $M$ instead of the best competing model at time $t$.

The \textit{Edge Ratio} for model $M$ is defined as
\[
    \text{EdgeRatio}(M)
    =
    \frac{\sum_{t}^T e_{M,t}^{+}}
         {\sum_{t}^T  e_{M,t}^{-}} \times (M-1) \enskip .
\]
The multiplicative factor $(M-1)$ corrects for the mechanical effect of the size of the model set. Under a null in which all $M$ models are equally informative, the probability that any given model attains the frontier at a given time is $1/M$. With equal expected magnitudes, the unscaled ratio $\sum e^+/\sum e^-$ therefore converges to the odds of winning versus losing, $1/(M-1)$, as the number of competing models increases. Scaling by $(M-1)$ normalizes this effect and yields a baseline value of approximately one under the null of no unique predictive advantage.

From a decision-theoretic perspective, the Edge Ratio can be interpreted as the ratio of \textit{avoided regret} (when model $M$ wins) to \textit{incurred regret} (when it fails). The numerator captures what a forecaster would have lost by not choosing $M$ when it was the best option; the denominator captures what was lost by choosing $M$ when a better alternative existed. A high Edge Ratio thus indicates that the benefits of including $M$ in the forecaster's toolkit outweigh the costs of its occasional failures---the model delivers a persistent and economically meaningful edge relative to the forecasting frontier. By contrast, a model with strong average performance but a low Edge Ratio provides improvements that are largely replicable by other available methods. In the limit, a model that never attains the frontier (for instance, one that consistently finishes second) receives an Edge Ratio of exactly zero, since $\sum e^+_t = 0$.

\paragraph{Pool Sensitivity.} Unlike benchmark-relative metrics such as the Sharpe or Sortino ratio, which depend only on a fixed benchmark model, the Edge Ratio is inherently a function of the model pool: adding or removing competitors alters the frontier $L_t^\star$ and thereby changes every model's edge. The $(M-1)$ scaling factor normalizes for pool \emph{size}---ensuring that the baseline value remains near one regardless of the number of models---but it does not neutralize changes in pool \emph{composition}. Adding a volatile, distinctive model can shift the frontier at specific time periods, potentially altering rankings; conversely, adding near-duplicates will leave the frontier largely unchanged while increasing $M$, an effect absorbed by the scaling. This is analogous to a well-known feature of benchmark-relative metrics: the Sharpe ratio of a model changes if one switches from an AR to the SPF as benchmark. The Edge Ratio simply replaces a fixed benchmark with a moving, pool-dependent one. As such, Edge Ratios are most informative when the model pool is held fixed across comparisons, as in the applications presented here.





\subsection{Meta-Analysis Statistics}
\label{sec:meta_logic}

Empirical forecasting studies typically evaluate models across many targets, horizons, and design choices. Beyond per-series, time-domain risk metrics based on $\{r_t\}$, it is therefore useful to assess performance stability at a more aggregated, \emph{meta} level.

Let targets be indexed by $v = 1, \dots, V$, forecasting horizons by $h = 1, \dots, H$, and evaluation designs by $s = 1, \dots, S$. For each combination $(v,h,s)$ and model $M$, let $P_{v,h,s}^{M}$ denote a scalar performance metric (e.g., MSE, RMSE, or MAE), with $P_{v,h,s}^{B}$ the corresponding benchmark value. I define a percentage ``return'' relative to the benchmark as
\[
    R_{v,h,s}^{M}
    =
    \frac{P_{v,h,s}^{B} - P_{v,h,s}^{M}}{P_{v,h,s}^{B}} \times 100.
\]
For example, an RMSE ratio of $0.90$ relative to the benchmark corresponds to a return of $10\%$. The collection $\{R_{v,h,s}^{M}\}$ forms a \textit{cross-sectional return distribution} for model $M$ over the design space. Stacking indices $(v,h,s)$ into a single index $i = 1, \dots, N$, with $N = V \times H \times S$, the cross-sectional mean and dispersion are
\[
    \bar{R}^{M} = \frac{1}{N} \sum_{i=1}^{N} R_i^{M},
    \qquad
    (s_R^{M})^2 = \frac{1}{N-1} \sum_{i=1}^{N} (R_i^{M} - \bar{R}^{M})^2.
\]
The resulting \textit{cross-sectional Sharpe-type index} is
\[
    \text{Sharpe}_{\text{meta}}^{M}
    =
    \frac{\bar{R}^{M}}{s_R^{M}}.
\]
Analogously, $\text{Sortino}_{\text{meta}}^{M}$,  $\Omega_{\text{meta}}^{M}$, and $\text{Edge}_{\text{meta}}^{M}$ are computed by applying the same downside–upside decompositions to the cross-sectional distribution $\{R_i^{M}\}$. These statistics measure how reliably a model outperforms the benchmark \emph{across forecasting problems}, rather than over time within a single problem.  From a model development perspective, this distinction is important. A model with a strong meta-level Sortino or Omega ratio is one that tends to outperform the benchmark in many settings, {while avoiding catastrophic failures when it does not}. Meta-analysis statistics therefore complement time-series risk measures by characterizing the robustness of model improvements across targets, horizons, and evaluation designs.

\section{Application 1 : Predictive Personalities \& Macro Forecasting}\label{sec:showdown}

 I conduct a comprehensive macroeconomic forecasting exercise comparing classical econometric models, modern machine learning algorithms, a transformer-based foundation model, and the Survey of Professional Forecasters. The goal is not merely to rank models by average RMSE, but to characterize their \textit{predictive personalities}---the distinctive patterns of strengths, weaknesses, and risk profiles that determine their value in a forecasting portfolio.

\paragraph{Data and Forecasting Setup.} I forecast 4 U.S. macroeconomic variables at the quarterly frequency: headline CPI inflation, GDP growth, the unemployment rate, and housing starts (log growth rate). This set covers variables where the SPF provides benchmark forecasts, enabling direct comparison between ML models and professional judgment. Predictors are drawn from the FRED-QD database \citep{mccracken2020fred}, transformed to induce stationarity, and augmented with 4 lags plus moving averages of order 2, 4, and 8 following the MARX transformation \citep{medeiros2021forecasting}. All series are standardized to zero mean and unit variance over the training sample. I estimate direct forecasts at horizons $h = 1, 2, 4$ quarters ahead, with separate models for each horizon.

The out-of-sample evaluation proceeds on an expanding window across two distinct periods, with models re-estimated every 8 quarters (2 years). Panel A (2007Q2–2019Q4) covers the Great Financial Crisis, the subsequent recovery, and the pre-pandemic expansion, with training data beginning in 1961Q2. Panel B (2021Q1–2024Q2) covers the post-pandemic recovery and the 2021–2022 inflation surge; I exclude 2020 to avoid contaminating estimates with the COVID shock. Comparing performance across these regimes—one characterized by low inflation and gradual recovery, the other by unprecedented volatility and rapid policy shifts—allows one to assess whether ML gains are stable or driven by specific episodes.
\paragraph{Models.} The suite of models is an expanded universe from \citet{dual}, which itself follows in the footsteps of a tradition of various ML models in macroeconomic forecasting.

\begin{enumerate}[leftmargin=6em, labelwidth=5em, labelsep=0.5em, align=left]
    \item[\texttt{\fontfamily{phv}\selectfont \textbf{\phantom{Tab}AR(4)}:}] Fourth-order autoregression on the target variable. The benchmark that any serious forecaster must beat.

    \item[\texttt{\fontfamily{phv}\selectfont \textbf{\phantom{Ta}FAAR}:}] Factor-Augmented AR \citep{stock2002forecasting}. Augments the AR(4) with four principal components extracted from the predictor panel, exploiting cross-variable information while avoiding the curse of dimensionality.

    \item[\texttt{\fontfamily{phv}\selectfont \textbf{\phantom{TabP}RR}:}] Ridge Regression. High-dimensional linear prediction with $L_2$ penalty; regularization parameter $\lambda$ selected by cross-validation.

    \item[\texttt{\fontfamily{phv}\selectfont \textbf{\phantom{Tab}KRR}:}] Kernel Ridge Regression. Extends ridge to nonlinear settings via Gaussian and Laplacian kernels; bandwidth $\sigma$ and $\lambda$ cross-validated.

    \item[\texttt{\fontfamily{phv}\selectfont \textbf{\phantom{TabP}RF}:}] Random Forest \citep{breiman2001random}. Aggregates 500 trees grown on bootstrap samples with random feature subsets; 75\% subsampling, minimum node size 5.

    \item[\texttt{\fontfamily{phv}\selectfont \textbf{\phantom{Tab}LGB}:}] LightGBM \citep{ke2017lightgbm}. Sequential gradient boosting with histogram-based splits; learning rate, depth, and sampling fractions cross-validated with early stopping.

    \item[\texttt{\fontfamily{phv}\selectfont \textbf{\phantom{Ta}LGB+}:}] Hybrid LightGBM \citep{lgbplus}. At each boosting step, a tree-based and a linear update compete; the winner is selected via out-of-bag validation.

    \item[\texttt{\fontfamily{phv}\selectfont \textbf{LGB$^{\texttt{A}}$+}:}] The alternating variant of LGB+ \citep{lgbplus}. A more computationally economical version that alternates tree ensembles with linear corrections in a fixed pattern each boosting cycle.

    \item[\texttt{\fontfamily{phv}\selectfont \textbf{\phantom{TabP}NN}:}] Feed-forward neural network with three hidden layers (400 neurons each), ReLU activations, and dropout regularization (rate 0.2); trained via Adam optimizer with early stopping.

    \item[\texttt{\fontfamily{phv}\selectfont \textbf{\phantom{Tab}HNN}:}] Hemisphere Neural Network \citep{HNN}. A constrained neural architecture with four dedicated hemispheres (long-run expectations, short-run expectations, output gap, commodities) sharing a common feature core, enabling proactive volatility forecasting.

    \item[\texttt{\fontfamily{phv}\selectfont \textbf{TabPFN}:}] A transformer-based foundation model \citep{hollmann2022tabpfn} (TPFN hereafter) pre-trained on millions of synthetic tabular datasets. Unlike conventional ML that requires task-specific training, TabPFN performs in-context learning---ingesting training data as context and producing predictions in a single forward pass. A key advantage is that its training on purely synthetic data mitigates concerns about data leakage that can arise with foundation models trained on web-scraped corpora \citep[see][for evidence that LLM-based inflation forecasts suffer from look-ahead bias in practice]{alam2026chatmacro}. Whether patterns learned across synthetic problems transfer to macroeconomic forecasting, with its temporal dependencies and regime changes, is an open question.

    \item[\texttt{\fontfamily{phv}\selectfont \textbf{\phantom{Tab}SPF}:}] Survey of Professional Forecasters. The median aggregates predictions from dozens of professionals with access to proprietary models, real-time data, and qualitative judgment unavailable to statistical methods \citep{stark2010realistic,engelberg2022manskie}. 
\end{enumerate}



\subsection{Results}
\label{sec:results_app1}

I forecast a large number of variables across multiple horizons and evaluation designs. Complete results for all risk-adjusted metrics discussed above, along with classical performance measures such as RMSEs and mean absolute errors, are reported in the Appendix (Table~\ref{tab:gdp_h1} and all those that follow). To facilitate the excavation of such a large body of results, I begin by visualizing three cases of particular interest: GDP growth, unemployment, and two additional targets—inflation and housing starts—that allow for direct comparison with the Survey of Professional Forecasters. It is worth noting that the ``returns'' $r_t = L_t^B - L_t^M$ are defined in terms of squared errors, not root mean squared errors, since this is the natural loss differential that arises from comparing forecast accuracy.

\subsubsection{GDP Growth}

\begin{figure}[t!]
\centering
\caption{Risk-Adjusted Performance for GDP Growth, Pre-COVID}
\label{fig:gdp_barplot}
\includegraphics[width=1.01\linewidth]{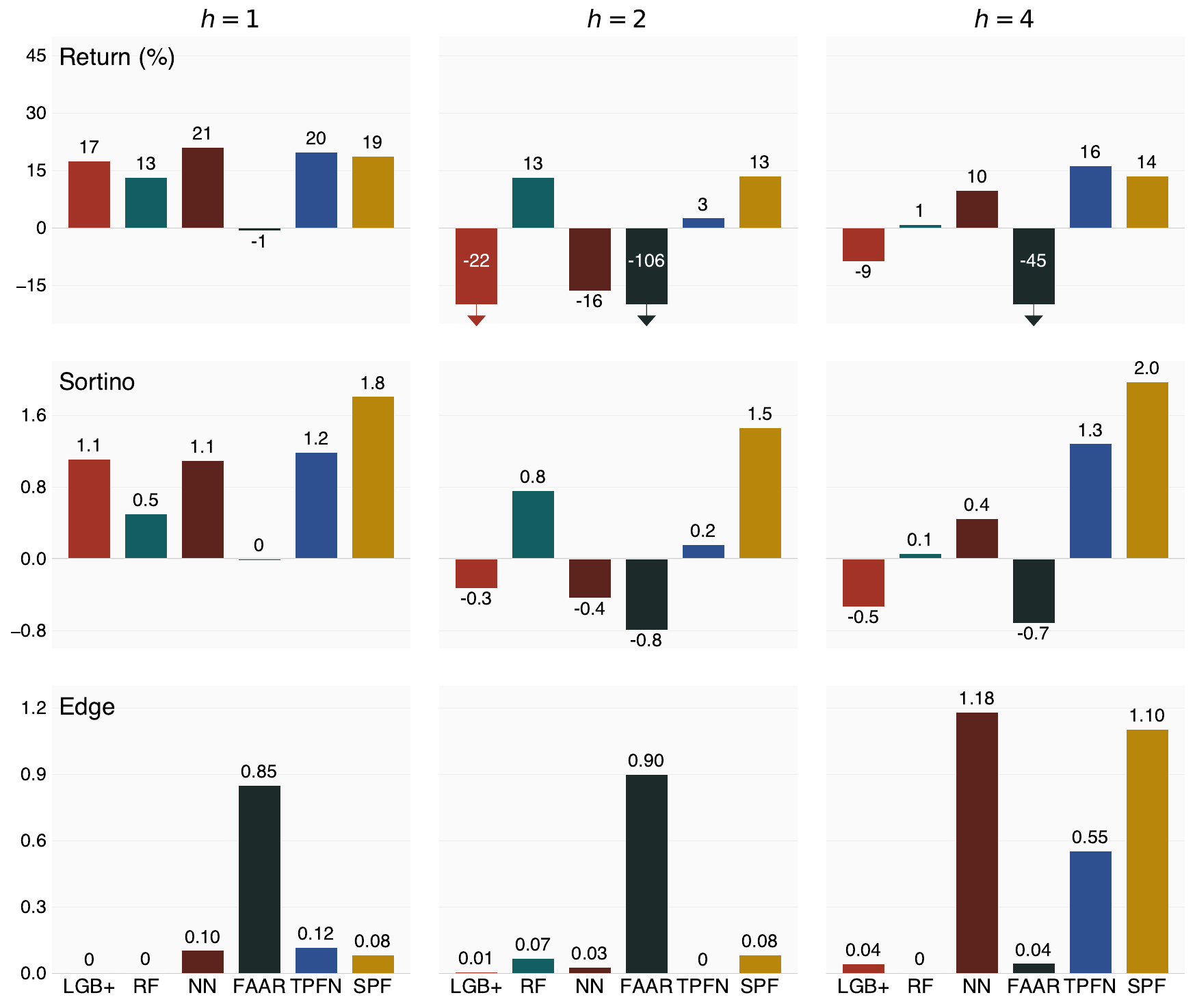}
\par\medskip
\parbox{\textwidth}{\footnotesize \textit{Notes:} Each row displays a different metric (Return, Sortino, Edge) across forecast horizons $h \in \{1, 2, 4\}$. Detailed results are reported in the Appendix.}
\end{figure}

Figure \ref{fig:gdp_barplot} reports results for GDP growth in the pre-COVID sample (2007--2019). At the short horizon ($h=1$), Kernel Ridge Regression delivers the highest average return, followed by neural networks and the foundation model, consistent with the ability of flexible models to exploit short-run nonlinearities. However, once downside risk is taken into account, the ranking shifts noticeably. The SPF's Sortino ratio (1.80) exceeds that of neural networks (1.09), consistent with low downside volatility---the SPF rarely underperforms the benchmark, even when its average return is moderate. KRR's risk-adjusted profile is even stronger (Sortino 2.85), combining high returns with tightly contained losses. At longer horizons ($h=2$ and $h=4$), the SPF's dominance becomes clear. At $h=2$, only RF and the SPF deliver positive average returns among the main competitors, while neural networks turn negative. What separates the SPF is lower downside risk: its Sortino ratios lead by a considerable margin, suggesting that its advantage in this sample lies more in stability than in aggressive gains.

\paragraph{Edge and Information Content.}
The Edge Ratio at $h=1$ highlights a complementary dimension of performance. The factor-augmented autoregression displays substantial edge (0.85), suggesting repeated episodes in which it delivers information unavailable to other models. This is consistent with a highly uneven predictive personality: the factor model struggles most of the time, but occasionally captures common dynamics that other approaches miss, yielding sharp but infrequent gains. The SPF's Edge Ratio, by contrast, is modest at $h=1$ (0.08), suggesting that its short-horizon strength lies in consistency rather than in unique informational content.

Interestingly, this pattern reverses at longer horizons: the SPF's Edge Ratio rises substantially at $h=4$ (1.10), suggesting that professional forecasters provide uniquely informative signals at longer horizons---consistent with the value of judgment and institutional knowledge for medium-term outlook assessment. The factor model also continues to exhibit episodic edge across horizons, reinforcing its role as a specialist rather than a consistently strong forecaster.

\paragraph{Omega Ratio and Maximum Drawdown.} The SPF leads on the Omega ratio, but Ridge Regression often appears essentially tied in terms of upside--downside balance. KRR, omitted from Figure \ref{fig:gdp_barplot}, performs particularly well on this metric and delivers one of the strongest overall Omega profiles for GDP growth. However, these gains are fundamentally conservative in nature. KRR exhibits extremely small Edge Ratios, indicating that it rarely attains the forecasting frontier. Its improvements are steady and reliable, but seldom decisive.

This conservative profile is mirrored in path-dependent risk measures. For GDP at $h=1$, maximum drawdowns are extremely small for both the SPF and KRR, suggesting that cumulative underperformance is tightly contained. By contrast, drawdowns are substantially larger for neural networks and especially for the foundation model, TPFN. While these models occasionally deliver meaningful upside, they do so at the cost of greater exposure to sustained losses. For risk-averse users, this distinction may be more consequential than differences in average accuracy.

\begin{figure}[t!]
\centering
\caption{Risk-Adjusted Performance for Unemployment Rate, Pre-COVID}
\label{fig:ur_barplot}
\includegraphics[width=1.01\linewidth]{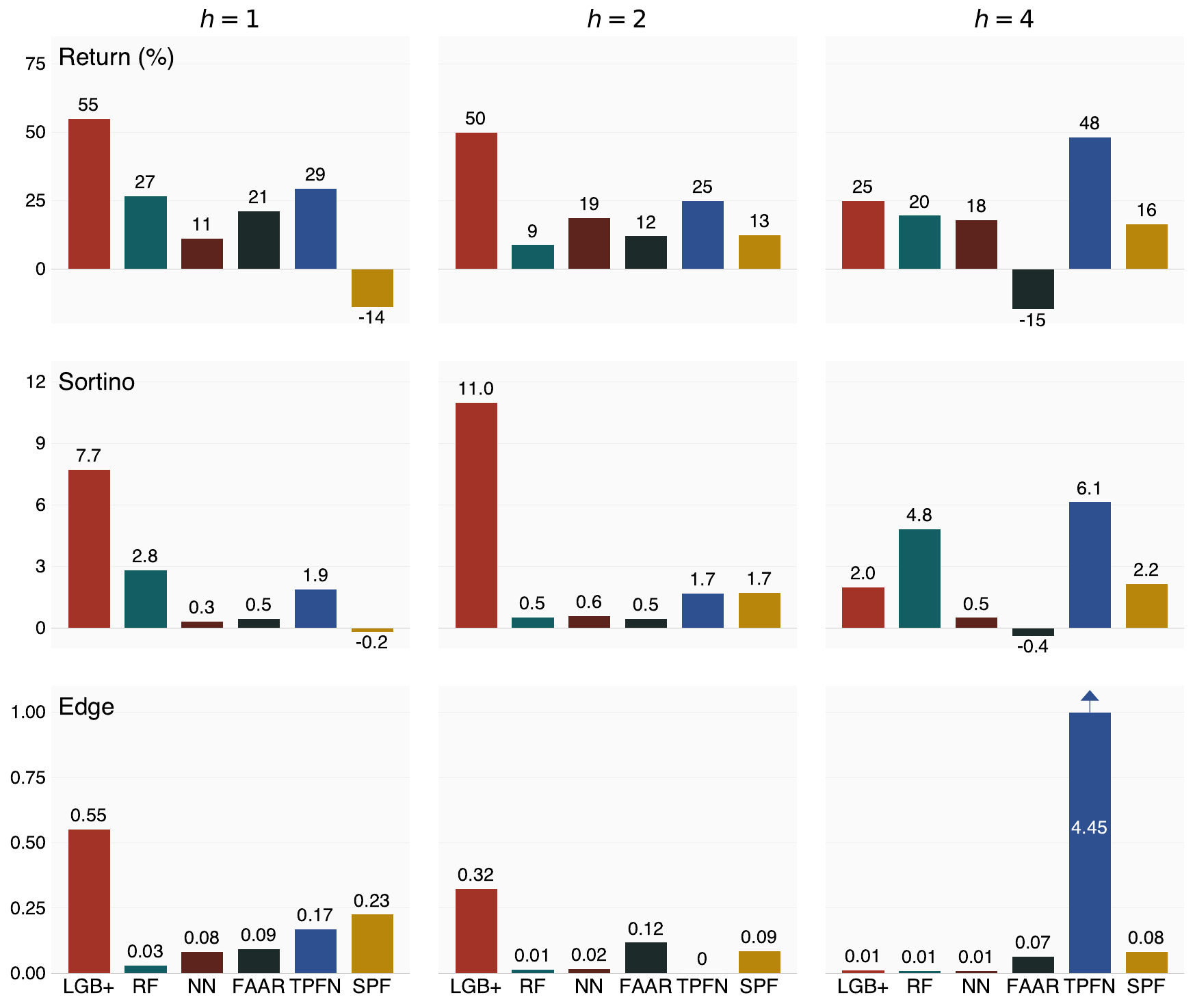}
\par\medskip
\parbox{\textwidth}{\footnotesize \textit{Notes:} Each row displays a different metric (Return, Sortino, Edge) across forecast horizons $h \in \{1, 2, 4\}$. Detailed results are reported in the Appendix.}
\end{figure}

\subsubsection{Unemployment Rate}
For unemployment (Figure~\ref{fig:ur_barplot}), a striking result is the strong performance of LGB+, the hybrid boosting model combining tree-based learners with linear basis functions. At the short horizon ($h=1$), LGB+ delivers high average returns while simultaneously exhibiting a strong Sortino ratio, suggesting that its gains are achieved with limited downside exposure. While it does not display the same predictive edge as neural networks, it appears to strike an unusually favorable balance between accuracy and risk, outperforming all competitors on a risk-adjusted basis at this horizon.

This performance contrasts sharply with that of the Survey of Professional Forecasters. In the pre-COVID sample, the SPF struggles to beat the autoregressive benchmark for unemployment at short horizons, both in terms of average returns and risk-adjusted metrics—a notable result given the SPF’s dominance for other macroeconomic aggregates such as GDP growth. 

LGB+ continues to lead at $h=2$, with an average return of 50\% and a Sortino of 11.0. TPFN emerges as a credible second (Return 25\%, Sortino 1.7), while neural networks (19\%) and the SPF (13\%) follow. At $h=4$, however, TPFN becomes the clear standout, posting the highest average return (48\%), Sortino ratio (6.1), and Edge Ratio (4.45) among all models. This may reflect the advantage of leveraging a large pretrained model at horizons where time series persistence and dynamics are harder to exploit. The SPF also improves at longer horizons: although its average gains remain modest (approximately 13\% at $h=2$ and 16\% at $h=4$), its substantially lower downside risk places its Sortino and Omega ratios broadly on par with several machine-learning models despite weaker raw returns.

Outside of the foundation model, the $h=4$ leaderboard in average returns is LGB+ (25\%), Random Forest (20\%), and neural networks (18\%), all in a similar range. Yet risk adjustment separates them: Random Forest, known as a robust predictive algorithm rarely prone to utter failure \citep{coulombe2025bag}, achieves a Sortino ratio (4.8) that approaches TPFN's (6.1), while the neural network's Sortino (0.5) is an order of magnitude lower. In other words, Random Forest delivers nearly the same risk-adjusted performance as the more sophisticated foundation model, while neural networks are blown out of the water once downside risk is accounted for.


\subsubsection{Inflation and Housing}
\begin{figure}[t!]
\centering
\caption{Risk-Adjusted Performance for Inflation and Housing Starts}
\label{fig:other_cases}
\includegraphics[width=1.01\linewidth]{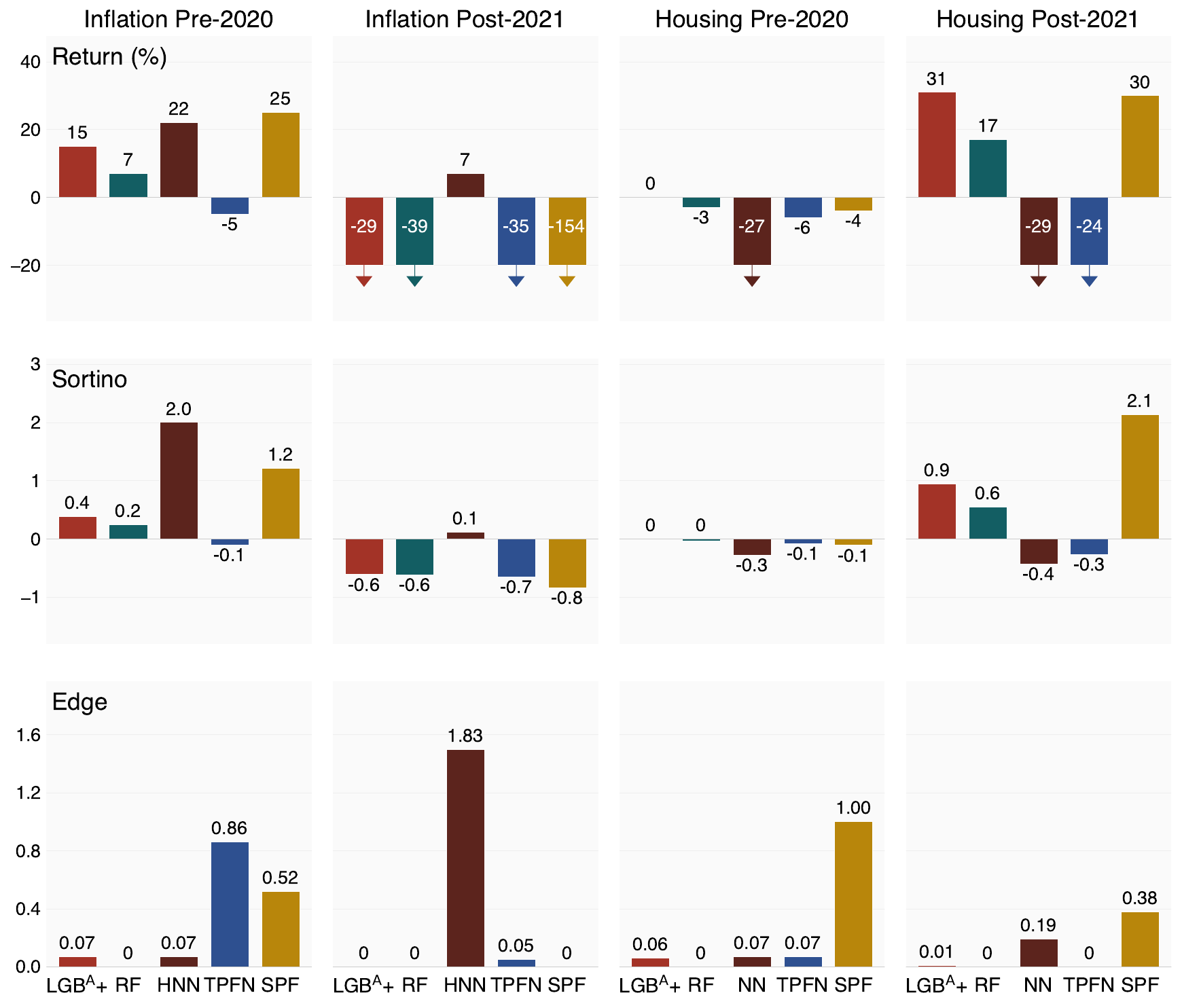}
\par\medskip
\parbox{\textwidth}{\footnotesize \textit{Notes:} The figure contrasts risk-adjusted performance for inflation and housing starts before and after the pandemic, at horizon $h=1$. Each row displays a different metric (Return, Sortino, Edge) across evaluation windows.}
\end{figure}

Inflation and housing starts (Figure~\ref{fig:other_cases}) allow for pre- and post-pandemic comparison. For quarterly inflation in the pre-pandemic period, the results are consistent with a well-known stylized fact: the SPF is extremely difficult to beat \citep{faust2013now,ang2007macro}. Across standard return measures, most statistical and machine learning models fail to improve meaningfully upon the SPF, a finding consistent with the broader literature documenting the resilience of professional inflation forecasts. Only HNN comes close to matching the SPF in terms of average performance, with LGB$^{\texttt{A}}$+ also delivering competitive returns.

Once downside risk is taken into account, HNN separates from the field. On a risk-adjusted basis, its Sortino ratio (2.0) exceeds the SPF's (1.2), indicating that HNN's gains come with contained downside exposure. The Edge Ratio provides additional nuance. The SPF exhibits the second-highest edge (0.52), consistent with the idea that professional forecasters occasionally exploit short-run information unavailable to purely data-driven models. TPFN achieves the highest Edge Ratio in this setting (0.86), suggesting that despite trailing the SPF in average returns, it occasionally delivers notable wins at the forecasting frontier.

The post-2021 period is dominated by the inflation surge and its aftermath, and forecasting performance deteriorates across nearly all models, including the SPF. The autoregressive benchmark, estimated on a long expanding window extending back to the 1970s, exhibits strong persistence and performs unusually well during this episode, effectively raising the bar for all competitors. Against this demanding benchmark, HNN is the only model retaining a positive average return. Moreover, HNN's Edge Ratio is by far the highest in the pool, indicating that it continues to reach the forecasting frontier when other models cannot. This is consistent with the findings of \cite{HNN}, where HNN's ability to extract nonlinear supervised summaries of leading indicators proves particularly valuable around inflation turning points. 

Housing starts tell a different story. In the pre-COVID period, no ML model meaningfully beats the autoregressive benchmark: LGB$^{\texttt{A}}$+ essentially breaks even (0\%), while Random Forest ($-3$\%) and neural networks ($-27$\%) fall short. The SPF also struggles ($-4$\%), and the SPF's Edge Ratio of 1.00 indicates that whenever any model does reach the frontier, it is the SPF. In the post-2021 period, LGB$^{\texttt{A}}$+ leads (Return 0.31), performing on par with the SPF (0.30). Random Forests also perform competitively, particularly in the post-2021 period. That said, the apparent strength of ML models in terms of raw returns partly evaporates once risk is accounted for. Although LGB$^{\texttt{A}}$+ matches the SPF in average post-2021 returns, adjusting for downside risk reveals a wide gap: the SPF's Sortino ratio (2.13) considerably exceeds that of LGB$^{\texttt{A}}$+ (0.95), indicating that the SPF achieves comparable gains with substantially less downside exposure. In this case, the SPF does not trade off upside for stability: it matches the best ML model in average returns while incurring substantially less downside risk. Post-COVID, the SPF's Edge Ratio (0.38) remains the highest in the pool, suggesting that it continues to reach the frontier more often than any ML competitor.

\section{Application 2 : Meta-Analyses}\label{sec:meta}

This section applies the meta-analysis statistics developed in Section~\ref{sec:meta_logic} to two distinct settings. First, I revisit the HNN forecasting exercise from \citet{GCFK}, assessing point and density forecast performance across multiple targets and horizons. Second, I analyze the M4 forecasting competition, where standardized evaluation across 48,000 monthly series provides an external benchmark. Together, these applications demonstrate that risk-adjusted metrics reveal patterns invisible to standard RMSE comparisons.

\subsection{Meta-Analysis 1 : Density Forecasting Evaluation}\label{sec:meta_hnn}

Table \ref{tab:meta_results} reports meta-analysis results based on the forecasting exercise of \citet{GCFK}. That paper adapts the HNN framework to density forecasting, extending it into a constrained neural architecture that jointly predicts conditional means and volatilities. A distinguishing feature of HNN is its capacity for proactive volatility forecasting: rather than relying solely on past prediction errors to update variance estimates—the reactive approach of traditional GARCH and stochastic volatility models—HNN can leverage leading indicators to anticipate heightened uncertainty before large forecast errors materialize. A central finding is that HNN not only performs well on average but does so reliably: unlike machine-learning competitors such as BART or DeepAR (see Appendix~\ref{app:gcfk} for model descriptions), it avoids episodes of severe underperformance in \textit{density} forecasting as measured by log scores.

The underlying empirical exercise is fairly extensive. Forecasts are evaluated for 5 quarterly target variables (GDP growth, unemployment, inflation, equity returns, and housing starts) at two forecasting horizons (1 and 4 quarters ahead), over two evaluation windows (2007Q1–2019Q4 and 2007Q1–2022Q4). Table \ref{tab:all_targets_rmse} reports full target-level RMSE and log-score results. While competing models occasionally dominate for specific targets or horizons, they also produce large negative log-score realizations in some configurations—what \citet{GCFK} term catastrophic density forecast failures. HNN, by contrast, delivers stable performance across this design space.

\begin{table}[t!]
  \small\centering
  \caption{\normalsize Model Performance: RMSE and Log Score Analysis \vspace*{-0.15cm}}
  \label{tab:meta_results}

  {\sffamily
  \begin{threeparttable}
    \setlength{\tabcolsep}{0.33em}
    \setlength\extrarowheight{7pt}

    \begin{tabular}{l rrrrrr c rrrrrr}
      \toprule \toprule
      & \multicolumn{6}{c}{\textbf{RMSE}} && \multicolumn{6}{c}{\textbf{Log Score}} \\
      \cmidrule(lr){2-7} \cmidrule(lr){9-14}
      \textbf{Model} & Return & Vol & Sharpe & Sortino & Omega & Edge
                     && Return & Vol & Sharpe & Sortino & Omega & Edge \\
      \midrule

      HNN
      & \textbf{\color{ForestGreen}11.8} & 13.9 & 0.85 & \textbf{\color{ForestGreen}1.62} & 2.1 & \textbf{\color{ForestGreen}1.97}
      && \textbf{\color{ForestGreen}17.1} & 35.8 & \textbf{\color{ForestGreen}0.48} & \textbf{\color{ForestGreen}0.92} & \textbf{\color{ForestGreen}2.4} & \textbf{\color{ForestGreen}2.59} \\

      \rowcolor{gray!7}
      BART
      & 10.6 & 11.0 & \textbf{\color{ForestGreen}0.97} & \textbf{\color{ForestGreen}1.62} & \textbf{\color{ForestGreen}2.4} & 0.81
      && -47.7 & 115.3 & -0.41 & -0.30 & 0.4 & 1.34 \\

      DeepAR
      & 3.8 & \textbf{\color{ForestGreen}10.1} & 0.38 & 0.91 & 1.7 & 0.38
      && -99.7 & 149.7 & -0.67 & -0.52 & 0.0 & 2.10 \\

      \rowcolor{gray!7}
      BLR
      & 5.2 & 13.8 & 0.38 & 0.54 & 1.5 & 0.58
      && 3.4 & \textbf{\color{ForestGreen}21.6} & 0.16 & 0.28 & 1.4 & 0.57 \\

      NN\textsubscript{G}
      & -1.7 & 21.6 & -0.08 & -0.08 & 0.9 & 0.69
      && 5.9 & 26.0 & 0.23 & 0.36 & 1.4 & 1.04 \\

      \rowcolor{gray!7}
      NN\textsubscript{SV}
      & -1.7 & 21.6 & -0.08 & -0.08 & 0.9 & 0.69
      && 1.9 & 28.5 & 0.07 & 0.09 & 1.1 & 0.50 \\


      \bottomrule \bottomrule
    \end{tabular}

    \begin{tablenotes}[para,flushleft]
      \scriptsize
      \textit{Notes}: Return = mean percentage improvement over AR benchmark; Vol = standard deviation across target–horizon–period combinations; Sharpe = Return/Vol; Sortino = Return/Downside deviation; Omega = Upside/Downside ratio; Edge = Edge Ratio (scaled by $M-1$; null expectation = 1). Bold green indicates best in column.
    \end{tablenotes}

  \end{threeparttable}
  }
\end{table}

Table \ref{tab:meta_results} revisits this qualitative observation more formally, looking at ``returns'' in point forecasts (RMSE) and in density forecasts (Log Score). Focusing first on point forecast performance (RMSE panel), HNN delivers a slightly larger average improvement over the AR benchmark than BART, although BART exhibits lower performance volatility. As a result, BART attains a marginally higher Sharpe ratio. When attention is restricted to downside risk, however, HNN and BART display identical Sortino ratios, indicating comparable exposure to adverse realizations. Using the more flexible Omega ratio suggests a modest advantage for BART in terms of overall risk profile, although both models clearly belong to the same performance cluster. This pattern highlights the robustness of tree-based methods for point forecasting tasks.

At the same time, the Edge Ratio is substantially higher for HNN, indicating that when HNN outperforms competing models, it does so by a wider margin than the margin by which it underperforms when it fails to be the best. This asymmetry provides quantitative support for the proactivity arguments emphasized in \citet{GCFK}: HNN’s gains tend to be decisive rather than marginal. Although BART exhibits lower raw volatility in point forecast performance, its lower average return limits its ranking on risk-adjusted RMSE criteria.

The contrast is sharper in the density forecast evaluation (log-score panel), where HNN achieves a decisive improvement over all competing models. While linear regression and neural-network specifications with reactive volatility dynamics exhibit lower volatility of log-score improvements, their average performance gains are substantially smaller. HNN’s large mean improvement, combined with contained volatility and particularly low downside exposure, results in superior Sortino, Omega, and Edge ratios. In addition, its high Edge Ratio indicates that HNN consistently distances competing models in density forecasting, reinforcing the evidence that its advantage lies not only in average performance but also in the reliability and distinctiveness of its gains.

\subsection{Meta-Analysis 2 : The M4 Forecasting Competition}
\label{sec:m4_external}

Next, I apply the same risk-adjusted framework to the M4 forecasting competition \citep{makridakis2018m4,makridakis2020m4}, one of the largest and most widely used benchmarks for forecast evaluation. The M4 competition comprises 100{,}000 time series spanning multiple frequencies and domains, with forecasts evaluated using standardized loss functions and a common evaluation protocol. I focus on the 48{,}000 monthly series, which are assessed using MASE and OWA, the latter combining relative MASE and sMAPE into a single summary metric. As in Section~\ref{sec:meta_hnn}, this is a meta-analysis: the ``returns'' are not raw forecast errors from individual time series, but rather the distribution of reported performance metrics (MASE and OWA improvements) across the large cross-section of monthly series. This makes it possible to assess not just which method wins on average, but how robust that advantage is across the design space.

\begin{table}[t!]
  \small\centering
  \caption{\normalsize M4 Monthly: Meta-Metrics for Forecast Evaluation \vspace*{-0.3cm}}
  \label{tab:m4_meta_metrics}
  {\sffamily
  \begin{threeparttable}
    \setlength{\tabcolsep}{0.4em}\setlength\extrarowheight{5pt}
    \begin{tabular}{l rrrrrr c rrrrrr}
      \toprule \toprule
      & \multicolumn{6}{c}{\textbf{MASE}} && \multicolumn{6}{c}{\textbf{OWA}} \\
      \cmidrule(lr){2-7} \cmidrule(lr){9-14}
      \textbf{Model} & Return & Vol & Sharpe & Sortino & Omega & Edge
                     && Return & Vol & Sharpe & Sortino & Omega & Edge \\
      \midrule
      118 & \textbf{\color{ForestGreen}11.6} & 71.3 & \textbf{\color{ForestGreen}0.16} & \textbf{\color{ForestGreen}0.20} & \textbf{\color{ForestGreen}1.56} & \textbf{\color{ForestGreen}0.51}
          && 15.7 & \textbf{\color{ForestGreen}42.3} & 0.37 & 0.58 & 2.20 & \textbf{\color{ForestGreen}0.50} \\
      \rowcolor{gray!7}
      245 & 10.7 & \textbf{\color{ForestGreen}71.2} & 0.15 & 0.18 & 1.50 & 0.23
          && 16.4 & 46.8 & 0.35 & 0.52 & 2.32 & 0.19 \\
      72 & 10.5 & 72.6 & 0.14 & 0.18 & 1.48 & 0.04
          && 15.8 & 44.1 & 0.36 & 0.55 & 2.23 & 0.03 \\
      \rowcolor{gray!7}
      237 & 9.5 & 73.4 & 0.13 & 0.16 & 1.43 & 0.04
          && \textbf{\color{ForestGreen}18.7} & 44.6 & \textbf{\color{ForestGreen}0.42} & \textbf{\color{ForestGreen}0.64} & \textbf{\color{ForestGreen}2.98} & 0.05 \\
      69 & 9.3 & 73.2 & 0.13 & 0.15 & 1.41 & 0.02
          && 14.9 & 47.3 & 0.32 & 0.46 & 2.11 & 0.01 \\
      \rowcolor{gray!7}
      36 & 8.7 & 73.5 & 0.12 & 0.14 & 1.38 & 0.02
          && 15.4 & 49.2 & 0.31 & 0.45 & 2.22 & 0.02 \\
      132 & 7.9 & 75.5 & 0.10 & 0.13 & 1.34 & 0.38
          && 12.2 & 54.5 & 0.22 & 0.30 & 1.76 & 0.27 \\
      \rowcolor{gray!7}
      235 & 7.6 & 75.0 & 0.10 & 0.12 & 1.32 & 0.41
          && 15.4 & 49.9 & 0.31 & 0.43 & 2.32 & 0.37 \\
      78 & 7.1 & 74.1 & 0.10 & 0.12 & 1.30 & 0.19
          && 13.2 & 43.0 & 0.31 & 0.45 & 1.98 & 0.22 \\
      \rowcolor{gray!7}
      ARIMA & 7.0 & 75.6 & 0.09 & 0.11 & 1.29 & 0.39
             && 8.2 & 59.0 & 0.14 & 0.18 & 1.42 & 0.26 \\
      238 & 6.9 & 74.4 & 0.09 & 0.11 & 1.29 & 0.33
          && 15.5 & 45.9 & 0.34 & 0.49 & 2.37 & 0.27 \\
      \rowcolor{gray!7}
      239 & 6.5 & 74.9 & 0.09 & 0.10 & 1.27 & 0.28
          && 14.4 & 45.6 & 0.32 & 0.46 & 2.19 & 0.27 \\
      39 & 5.7 & 74.5 & 0.08 & 0.09 & 1.23 & 0.05
         && 9.8 & 48.3 & 0.20 & 0.28 & 1.59 & 0.05 \\
      \rowcolor{gray!7}
      ETS & 5.2 & 77.4 & 0.07 & 0.08 & 1.21 & 0.10
          && 10.8 & 57.2 & 0.19 & 0.24 & 1.67 & 0.07 \\
      104 & 4.8 & 77.7 & 0.06 & 0.07 & 1.19 & 0.39
          && 9.2 & 57.9 & 0.16 & 0.20 & 1.53 & 0.24 \\
      \bottomrule \bottomrule
    \end{tabular}
    \begin{tablenotes}[para,flushleft]
      \scriptsize
      \textit{Notes}: Return = percentage improvement relative to Naive2 (positive = better). Vol = standard deviation across series. Sharpe = Return/Vol; Sortino = Return/Downside Vol; Omega = gains-to-losses ratio. Edge = Edge Ratio (scaled by $M-1$; null expectation = 1). Bold green indicates best in column.
    \end{tablenotes}
  \end{threeparttable}}
\end{table}

Table~\ref{tab:m4_meta_metrics} reports risk-adjusted meta-metrics for the monthly M4 series, using percentage improvements relative to the Naive2 benchmark (a random walk with multiplicative seasonal adjustment when seasonality is detected, which serves as the standard baseline in the M4 protocol) as returns. Right off the bat, it is apparent that models ranking highly in terms of average MASE or OWA improvements also tend to exhibit similar levels of performance dispersion across the large cross-section of monthly series, resulting in modest MASE-based Sharpe ratios (approximately 0.10--0.16). OWA-based Sharpe ratios are notably higher (up to 0.42), reflecting the tighter dispersion of the composite metric. This suggests substantial heterogeneity in realized performance across series, but little evidence of a strong trade-off between average performance and volatility. Incorporating performance variability therefore does not materially overturn conventional average-performance rankings in this environment.

Model~118, the ES-RNN hybrid of \cite{smyl2020hybrid} which won the M4 competition, emerges as the most robust performer under MASE-based evaluation, achieving the highest Sharpe, Sortino, Omega, and Edge ratios in that panel. Under OWA, Model~237 attains the highest average return, but Model~118 remains highly competitive in risk-adjusted terms, suggesting that differences across the two panels are driven primarily by the choice of loss function rather than by fundamentally different risk profiles.

The Edge Ratio is particularly informative in the M4 context. In practice, users rarely forecast all series simultaneously; instead, they typically focus on a subset of series and aim to deploy the best-performing model on a case-by-case basis. From this perspective, the Edge Ratio captures a practically relevant notion of value by identifying models that are more likely to deliver a decisive advantage when selected for individual forecasting tasks. Even in a highly competitive setting such as M4—where average performance differences are small and volatility-adjusted rankings largely mirror mean rankings—a higher Edge Ratio indicates a greater propensity to generate average gains on  a target  that meaningfully distance the model from close competitors, rather than marginal improvements that are easily matched.

Finally, Edge Ratios remain relatively low overall, reflecting the intensity of competition in the M4 setting: even the best-performing methods rarely provide unique improvements that cannot be closely matched by alternative approaches. Nonetheless, Model~118 consistently attains the highest Edge Ratio among the leading models, indicating that when it does outperform the competition, it tends to do so by a wider margin, while avoiding large regrets when it fails to be the best. 



\section{Conclusion}
\label{sec:conclusion}

Average accuracy and forecast reliability are not the same thing. This paper shows how to tell them apart by treating loss differentials as returns and applying risk-adjusted metrics from finance. The Sharpe ratio often reorders model rankings relative to RMSE; the Sortino ratio and maximum drawdown reveal hidden fragility in models that look competitive on average. The Edge Ratio identifies which models contribute genuinely unique information to the forecasting frontier.

Applying this framework to a quarterly U.S. forecasting exercise, the Survey of Professional Forecasters is hard to beat not because it is the most accurate, but because it rarely fails badly. Among ML methods, tree-based models and HNN offer the most attractive risk–return profiles for specific targets. Moreover, foundation models get a credible first audition.

Beyond this empirical setting, the framework also applies across a design space rather than through time. Practitioners who adopt a method based on published evidence will inevitably apply it to new targets, horizons, or samples. The meta-analysis metrics developed here quantify the downside risk of doing so—revealing which methods are robust across the design space and which ones look good on average but fail badly in pockets.

Looking ahead, one direction for future research is to optimize forecasting models directly for risk-adjusted objectives. If the Sortino ratio is the metric that matters operationally, there is no reason to train on squared error and hope for favorable risk properties ex post. This is feasible within neural networks, and has a precedent in finance: \cite{cong2021alphaportfolio} use reinforcement learning to maximize the Sharpe ratio of portfolio returns directly. An analogous approach for forecasting could factor in institutional preferences at estimation time.


\clearpage
\bibliographystyle{apalike}
\bibliography{references}


\clearpage
\appendix

\section{Details on the GCFK Forecasting Exercise}\label{app:gcfk}

This appendix provides details on the data, models, and evaluation design underlying the meta-analysis in Table \ref{tab:meta_results}, which draws on \citet{GCFK}.

\vskip 0.25cm

{\noindent \sc \textbf{Data.}} The forecasting exercise uses the FRED-QD database of \citet{mccracken2020fred}, comprising 248 quarterly U.S.\ macroeconomic and financial series from 1960Q1 to 2022Q4. All variables except prices are transformed to achieve approximate stationarity following \citet{mccracken2020fred}; prices enter as log first differences (inflation rates). Predictors are standardized to zero mean and unit variance. Two lags of each variable are included, along with 100 linear trends to accommodate slow parameter drift. Missing values at the start of the sample are imputed via the EM algorithm.

\vskip 0.25cm

{\noindent \sc \textbf{Target Variables and Forecast Horizons.}} Forecasts are produced for five target variables: real GDP growth, change in the unemployment rate, headline CPI inflation, housing starts growth, and S\&P 500 returns. Each target is forecast at horizons $s \in \{1, 4\}$ quarters ahead using direct forecasting. The out-of-sample evaluation spans 2007Q1--2022Q4, with a secondary window ending in 2019Q4 (pre-pandemic). Neural network models are re-estimated every two years on an expanding window; other models are updated quarterly.

\vskip 0.25cm

{\noindent \sc \textbf{Models.}} The benchmark set includes:

\begin{enumerate}[leftmargin=5em, labelwidth=4em, labelsep=1em, align=left]
    \item[\texttt{ \fontfamily{phv}\selectfont \textbf{AR$_\text{G}$}:}] Standard autoregressive model with GARCH(1,1) errors. The conditional mean is an AR(2) and the variance follows the standard GARCH recursion.
    
    \item[\texttt{ \fontfamily{phv}\selectfont \textbf{AR$_\text{SV}$}:}] Autoregressive model with stochastic volatility. The log-variance follows an AR(1) process, estimated via MCMC using the algorithm of \citet{kastner2014ancillarity}.
    
    \item[\texttt{ \fontfamily{phv}\selectfont \textbf{\phantom{A}BLR}:}] A high-dimensional Bayesian linear regression with Normal-Gamma shrinkage priors \citep{griffin2010inference} and stochastic volatility. The hierarchical prior imposes both global and local shrinkage on the coefficient vector. Estimation proceeds via MCMC with 20,000 iterations and a burn-in of 10,000.
    
    \item[\texttt{ \fontfamily{phv}\selectfont \textbf{NN$_\text{SV}$}:}] Feed-forward neural network (two hidden layers, 400 neurons each, ReLU activation) trained by minimizing squared errors. Volatility is fitted in a second step on out-of-bag residuals using stochastic volatility.
    
    \item[\texttt{ \fontfamily{phv}\selectfont \textbf{NN$_\text{G}$}:}] Same architecture as NN$_\text{SV}$, but with GARCH(1,1) fitted to the out-of-bag residuals in the second step.
    
    \item[\texttt{ \fontfamily{phv}\selectfont \textbf{BART}:}] Bayesian additive regression trees \citep{BART} with 250 trees and stochastic volatility, following \citet{clark2022tail}. Tree structure priors and terminal node distributions follow standard recommendations.
    
    \item[\texttt{ \fontfamily{phv}\selectfont \textbf{DeepAR}:}] Amazon's autoregressive LSTM-based probabilistic forecasting model \citep{salinas2020deepar}. The architecture features two hidden layers of 400 LSTM cells with tanh activation, dropout rate of 0.2, and Gaussian likelihood. Training uses Adam optimizer with learning rate 0.001 over 20 epochs with patience of 5.
    
    \item[\texttt{ \fontfamily{phv}\selectfont \textbf{\phantom{D}HNN}:}] The Hemisphere Neural Network of \citet{GCFK}, with dedicated mean and variance hemispheres sharing a common input core. Each component has two hidden layers with 400 neurons. HNN is trained via maximum likelihood with three key modifications: (i) a volatility emphasis constraint that breaks mean/variance indeterminacy, (ii) blocked subsampling with $B=1000$ bootstrap samples, and (iii) an out-of-bag reality check for variance recalibration. ReLU activations are used throughout, with a softplus output for the variance hemisphere to ensure positivity.
\end{enumerate}

\vskip 0.25cm

{\noindent \sc \textbf{Evaluation Metrics.}} Point forecast accuracy is measured by RMSE relative to an AR(4) benchmark. Density forecast accuracy is measured by the average negative log predictive density ($\mathcal{L}$), where \textit{lower values indicate better density forecasts}. The benchmark for density evaluation is an AR model with stochastic volatility (AR$_\text{SV}$). Table~\ref{tab:all_targets_rmse} reports forecast performance in levels across all quarterly target variables for two evaluation periods: Panel A (2007Q2--2019Q4) and Panel B (2021Q1--2024Q2).

\begin{table}[t!]
  {\sffamily
\footnotesize\centering
\setlength{\tabcolsep}{0.35em}\setlength\extrarowheight{4.5pt}
\caption{\normalsize Forecast Performance of Quarterly Target Variables (RMSE and Log Score) \vspace*{-0.3cm}} \label{tab:all_targets_rmse}
\begin{threeparttable}
\begin{tabular}{l rrrrrrr c rrrrrrr}
\toprule \toprule
& \multicolumn{7}{c}{2007Q1 -- 2019Q4} && \multicolumn{7}{c}{2007Q1 -- 2022Q4} \\
\cmidrule(lr){2-8} \cmidrule(lr){10-16}
& HNN & NN\textsubscript{SV} & NN\textsubscript{G} & DeepAR & BART & AR\textsubscript{TV} & BLR && HNN & NN\textsubscript{SV} & NN\textsubscript{G} & DeepAR & BART & AR\textsubscript{TV} & BLR \\
\midrule
\rowcolor{gray!10} \multicolumn{16}{l}{\textbf{GDP ($s=1$)}} \\
RMSE & \textbf{0.83} & 0.93 & 0.93 & 0.92 & 0.86 & 1.01 & 0.89 && \textbf{0.85} & 0.96 & 0.96 & 0.93 & 0.92 & 1.00 & 0.94 \\
$\mathcal{L}$ & \textbf{-3.93} & -3.82 & -3.80 & -3.18 & -3.88 & -3.75 & -3.69 && \textbf{-3.87} & -3.70 & -3.63 & -3.23 & -3.71 & -3.69 & -3.63 \\
\rowcolor{gray!10} \multicolumn{16}{l}{\textbf{GDP ($s=4$)}} \\
RMSE & 0.90 & 0.88 & 0.88 & 1.07 & \textbf{0.88} & 0.99 & 0.91 && 0.85 & 1.46 & 1.46 & 0.99 & \textbf{0.81} & 0.98 & 0.95 \\
$\mathcal{L}$ & -3.70 & -3.54 & -3.52 & -2.83 & \textbf{-3.70} & -3.04 & -3.59 && \textbf{-3.61} & -3.33 & -3.36 & 1.27 & -3.55 & -3.05 & -3.51 \\
\rowcolor{gray!10} \multicolumn{16}{l}{\textbf{Unemployment ($s=1$)}} \\
RMSE & \textbf{0.73} & 0.78 & 0.78 & 0.93 & 0.87 & 1.00 & 0.75 && \textbf{0.82} & 0.96 & 0.96 & 0.90 & 0.96 & 1.04 & 0.91 \\
$\mathcal{L}$ & \textbf{-0.37} & -0.32 & -0.33 & 0.03 & 1.87 & -0.18 & -0.09 && \textbf{-0.24} & -0.11 & -0.09 & 0.10 & 1.85 & -0.04 & 0.05 \\
\rowcolor{gray!10} \multicolumn{16}{l}{\textbf{Unemployment ($s=4$)}} \\
RMSE & 0.74 & \textbf{0.69} & \textbf{0.69} & 0.85 & 0.75 & 0.97 & 0.82 && 0.70 & 2.20 & 2.20 & 0.73 & 0.71 & 0.88 & \textbf{0.70} \\
$\mathcal{L}$ & \textbf{-0.17} & 0.04 & 0.06 & 0.23 & 0.65 & 0.47 & 0.19 && \textbf{0.03} & 0.62 & 0.34 & 2.91 & 0.82 & 0.70 & 0.30 \\
\rowcolor{gray!10} \multicolumn{16}{l}{\textbf{Inflation ($s=1$)}} \\
RMSE & \textbf{0.94} & 0.95 & 0.95 & 1.02 & 1.07 & 1.11 & 1.05 && 1.14 & 1.17 & 1.17 & \textbf{0.93} & 0.96 & 1.00 & 1.23 \\
$\mathcal{L}$ & -3.63 & -3.74 & \textbf{-3.82} & -3.57 & -2.91 & -3.26 & -3.60 && -3.41 & -2.72 & -3.47 & \textbf{-3.52} & -1.30 & -3.32 & -3.33 \\
\rowcolor{gray!10} \multicolumn{16}{l}{\textbf{Inflation ($s=4$)}} \\
RMSE & \textbf{0.93} & 0.95 & 0.95 & 0.99 & 0.99 & 1.08 & 1.00 && 0.94 & 0.98 & 0.98 & 0.93 & \textbf{0.92} & 1.02 & 1.09 \\
$\mathcal{L}$ & -3.48 & -3.63 & \textbf{-3.66} & -3.14 & -3.19 & -3.45 & -3.60 && -3.30 & -3.39 & -3.40 & -1.27 & -2.99 & \textbf{-3.40} & -3.37 \\
\rowcolor{gray!10} \multicolumn{16}{l}{\textbf{S\&P 500 ($s=1$)}} \\
RMSE & 0.96 & 1.09 & 1.09 & 1.02 & \textbf{0.92} & 0.94 & 0.98 && 0.93 & 1.04 & 1.04 & 1.06 & \textbf{0.89} & 0.92 & 0.96 \\
$\mathcal{L}$ & \textbf{-1.55} & -1.24 & -1.27 & -1.34 & -1.28 & -1.35 & -1.25 && \textbf{-1.52} & -1.30 & -1.32 & -1.13 & -1.34 & -1.39 & -1.29 \\
\rowcolor{gray!10} \multicolumn{16}{l}{\textbf{S\&P 500 ($s=4$)}} \\
RMSE & 1.00 & 1.15 & 1.15 & 1.02 & \textbf{0.92} & 0.99 & 0.99 && 1.00 & 1.22 & 1.22 & 1.01 & \textbf{0.95} & 0.99 & 1.09 \\
$\mathcal{L}$ & -1.27 & -1.00 & -1.01 & 4.99 & \textbf{-1.36} & -1.16 & -1.25 && -1.27 & -0.96 & -0.99 & 3.80 & \textbf{-1.30} & -1.19 & -1.16 \\
\rowcolor{gray!10} \multicolumn{16}{l}{\textbf{Housing Starts ($s=1$)}} \\
RMSE & 0.99 & 1.01 & 1.01 & 1.06 & 0.96 & 0.99 & \textbf{0.96} && \textbf{0.86} & 0.87 & 0.87 & 0.97 & 0.93 & 1.00 & 0.99 \\
$\mathcal{L}$ & -1.14 & -1.08 & -1.08 & 0.07 & -0.98 & -1.15 & \textbf{-1.16} && -1.07 & -0.97 & -0.93 & -0.05 & -0.67 & \textbf{-1.15} & -0.92 \\
\rowcolor{gray!10} \multicolumn{16}{l}{\textbf{Housing Starts ($s=4$)}} \\
RMSE & 1.03 & 1.09 & 1.09 & 1.05 & 0.98 & \textbf{0.96} & 1.03 && 1.01 & 1.13 & 1.13 & 1.04 & 0.99 & \textbf{0.96} & 1.07 \\
$\mathcal{L}$ & -0.88 & -0.95 & -0.99 & 0.22 & \textbf{-1.10} & -1.05 & -1.02 && -0.66 & -0.51 & -0.66 & 0.35 & -0.75 & \textbf{-1.06} & -0.83 \\
\bottomrule \bottomrule
\end{tabular}
\begin{tablenotes}[para,flushleft]
\scriptsize
\textit{Notes}: RMSE relative to AR with constant variance. $\mathcal{L}$ = log score. Bold = best model per row/period. For GDP and unemployment, 2020 excluded. AR\textsubscript{TV} = best AR specification (SV or G).
\end{tablenotes}
\end{threeparttable}
}
\end{table}






\begin{landscape}
\begin{table}[t!]
  \centering
  \caption{\normalsize GDP ($h=1$)}
  \vspace*{-0.65em}
  \label{tab:gdp_h1}

  {\fontfamily{phv}\selectfont
  \resizebox{\linewidth}{!}{%
  \scriptsize
  \setlength{\tabcolsep}{0.3em}
  \renewcommand{\arraystretch}{1.65}
  \begin{tabular}{l r r r r r r r r r r c r r r r r r r r r r}
    \toprule

    & \multicolumn{10}{c}{\textcolor{dpd2}{\textbf{2007Q2 -- 2019Q4}}}
    && \multicolumn{10}{c}{\textcolor{dpd2}{\textbf{2022Q1 -- 2025Q1}}}
    \\
    \cmidrule(lr){2-11}
    \cmidrule(lr){13-22}
    & FAAR & RF & LGB & LGB+ & LGB$^{\texttt{A}}+$ & KRR & NN & RR & SPF & TPFN 
    && FAAR & RF & LGB & LGB+ & LGB$^{\texttt{A}}+$ & KRR & NN & RR & SPF & TPFN \\

    \midrule
    \addlinespace[0.5em]
    \multicolumn{22}{l}{\textcolor{dpd2}{\textbf{Panel A: Squared Error}}} \\
    \addlinespace[0.3em]

      Return
      & -0.01 & 0.13 & -0.05 & 0.17 & 0.17 & {\color{ForestGreen}\textbf{0.29}} & {\textbf{0.21}} & 0.17 & 0.19 & 0.20 &  & -1.21 & -0.12 & -0.96 & -0.14 & -0.29 & -0.23 & -0.31 & {\color{ForestGreen}\textbf{0.11}} & -0.45 & {\textbf{-0.03}} \\
      \rowcolor{RowAlt}
      Sharpe
      & -0.01 & 0.24 & -0.05 & 0.29 & 0.27 & {\textbf{0.44}} & 0.30 & 0.39 & {\color{ForestGreen}\textbf{0.47}} & 0.35 &  & -0.83 & -0.27 & -0.95 & -0.66 & -0.56 & -0.37 & -0.40 & {\color{ForestGreen}\textbf{0.31}} & -0.73 & {\textbf{-0.07}} \\
      Sortino
      & -0.02 & 0.50 & -0.09 & 1.11 & 0.74 & {\color{ForestGreen}\textbf{2.85}} & 1.09 & 1.01 & {\textbf{1.80}} & 1.18 &  & -0.82 & -0.38 & -0.95 & -0.75 & -0.64 & -0.43 & -0.44 & {\color{ForestGreen}\textbf{0.62}} & -0.79 & {\textbf{-0.09}} \\
      \rowcolor{RowAlt}
      Omega
      & 0.98 & 1.69 & 0.89 & 2.13 & 1.87 & {\color{ForestGreen}\textbf{4.03}} & 2.22 & 2.61 & {\textbf{3.78}} & 2.40 &  & 0.19 & 0.69 & 0.25 & 0.41 & 0.45 & 0.57 & 0.50 & {\color{ForestGreen}\textbf{1.57}} & 0.38 & {\textbf{0.90}} \\
      MaxDD
      & -13.63 & -0.49 & -12.20 & -0.34 & -0.61 & {\color{ForestGreen}\textbf{-0.11}} & -0.22 & -0.24 & {\textbf{-0.13}} & -0.26 &  & -17.99 & -4.21 & -14.92 & {\textbf{-1.45}} & -5.47 & -6.20 & -7.81 & {\color{ForestGreen}\textbf{-0.67}} & -8.06 & -3.46 \\
      \rowcolor{RowAlt}
      Edge
      & {\textbf{0.85}} & 0.00 & {\color{ForestGreen}\textbf{0.93}} & 0.00 & 0.17 & 0.02 & 0.10 & 0.09 & 0.08 & 0.12 &  & 0.05 & 0.00 & {\textbf{0.11}} & 0.00 & 0.00 & {\color{ForestGreen}\textbf{0.12}} & 0.00 & 0.00 & 0.08 & 0.05 \\

    \addlinespace[0.5em]
    \midrule
    \addlinespace[0.5em]
    \multicolumn{22}{l}{\textcolor{dpd2}{\textbf{Panel B: Absolute Error}}} \\
    \addlinespace[0.3em]

      Return
      & -0.12 & 0.01 & -0.12 & 0.02 & -0.01 & {\textbf{0.11}} & 0.04 & {\color{ForestGreen}\textbf{0.12}} & 0.07 & 0.04 &  & -0.63 & -0.23 & -0.66 & -0.13 & -0.27 & -0.29 & -0.31 & {\textbf{-0.10}} & -0.34 & {\color{ForestGreen}\textbf{-0.09}} \\
      \rowcolor{RowAlt}
      Sharpe
      & -0.25 & 0.05 & -0.28 & 0.08 & -0.03 & 0.40 & 0.13 & {\color{ForestGreen}\textbf{0.60}} & {\textbf{0.42}} & 0.16 &  & -0.92 & -0.61 & -1.17 & -0.58 & -0.64 & -0.60 & -0.59 & {\textbf{-0.34}} & -0.68 & {\color{ForestGreen}\textbf{-0.22}} \\
      Sortino
      & -0.32 & 0.07 & -0.34 & 0.13 & -0.04 & {\textbf{0.88}} & 0.23 & {\color{ForestGreen}\textbf{1.25}} & 0.76 & 0.24 &  & -0.89 & -0.69 & -1.10 & -0.66 & -0.73 & -0.65 & -0.62 & {\textbf{-0.41}} & -0.80 & {\color{ForestGreen}\textbf{-0.27}} \\
      \rowcolor{RowAlt}
      Omega
      & 0.72 & 1.07 & 0.66 & 1.12 & 0.96 & 1.81 & 1.23 & {\color{ForestGreen}\textbf{2.27}} & {\textbf{1.84}} & 1.24 &  & 0.20 & 0.45 & 0.21 & 0.45 & 0.44 & 0.43 & 0.39 & {\textbf{0.64}} & 0.45 & {\color{ForestGreen}\textbf{0.74}} \\
      MaxDD
      & -10.57 & -3.48 & -6.95 & -0.69 & -3.39 & {\textbf{-0.26}} & -0.52 & {\color{ForestGreen}\textbf{-0.23}} & -0.27 & -0.90 &  & -8.91 & -4.62 & -9.41 & {\color{ForestGreen}\textbf{-2.24}} & -5.01 & -5.64 & -6.09 & {\textbf{-3.28}} & -6.31 & -3.80 \\
      \rowcolor{RowAlt}
      Edge
      & {\color{ForestGreen}\textbf{1.10}} & 0.05 & {\textbf{0.47}} & 0.02 & 0.12 & 0.12 & 0.27 & 0.42 & 0.16 & 0.23 &  & 0.22 & 0.00 & 0.05 & 0.00 & 0.00 & 0.13 & 0.00 & 0.00 & {\color{ForestGreen}\textbf{0.34}} & {\textbf{0.29}} \\

    \addlinespace[0.5em]
    \midrule
    \addlinespace[0.5em]
    \multicolumn{22}{l}{\textcolor{dpd2}{\textbf{Panel C: Classical Forecast Accuracy}}} \\
    \addlinespace[0.3em]

      RMSE
      & 1.00 & 0.93 & 1.02 & 0.91 & 0.91 & {\color{ForestGreen}\textbf{0.84}} & {\textbf{0.89}} & 0.91 & 0.90 & 0.90 &  & 1.49 & 1.06 & 1.40 & 1.07 & 1.13 & 1.11 & 1.14 & {\color{ForestGreen}\textbf{0.95}} & 1.20 & {\textbf{1.02}} \\
      \rowcolor{RowAlt}
      MAE
      & 1.12 & 0.99 & 1.12 & 0.98 & 1.01 & {\textbf{0.89}} & 0.96 & {\color{ForestGreen}\textbf{0.88}} & 0.93 & 0.96 &  & 1.63 & 1.23 & 1.66 & 1.13 & 1.27 & 1.29 & 1.31 & {\textbf{1.10}} & 1.34 & {\color{ForestGreen}\textbf{1.09}} \\
      $\rho(1)$
      & 0.39 & 0.23 & {\color{ForestGreen}\textbf{0.07}} & 0.22 & {\textbf{0.11}} & 0.15 & 0.15 & 0.32 & 0.25 & 0.19 &  & 0.65 & 0.46 & 0.53 & 0.42 & 0.43 & 0.45 & 0.49 & {\textbf{0.39}} & 0.56 & {\color{ForestGreen}\textbf{0.36}} \\
      \rowcolor{RowAlt}
      DM $t$-stat
      & -0.03 & 1.04 & -0.23 & 1.03 & 0.92 & 1.41 & 0.95 & {\textbf{1.48}} & {\color{ForestGreen}\textbf{1.56}} & 1.24 &  & -1.26 & -0.50 & -1.84 & -1.19 & -1.01 & -0.55 & -0.59 & {\color{ForestGreen}\textbf{0.48}} & -1.32 & {\textbf{-0.13}} \\

    \addlinespace[0.3em]
    \bottomrule
  \end{tabular}%
  }

  \vspace{0.25em}
  \parbox{\linewidth}{\scriptsize
    \textit{Notes}: Panels A--B report risk-adjusted metrics; Panel C reports classical forecast accuracy metrics. $\rho(1)$ = first-order autocorrelation of errors.
    Best: \textcolor{ForestGreen}{\textbf{bold green}}; second-best: \textbf{bold}.
  }
  }

\end{table}
\end{landscape}


\begin{landscape}
\begin{table}[t!]
  \centering
  \caption{\normalsize GDP ($h=2$)}
  \vspace*{-0.65em}
  \label{tab:gdp_h2}

  {\fontfamily{phv}\selectfont
  \resizebox{\linewidth}{!}{%
  \scriptsize
  \setlength{\tabcolsep}{0.3em}
  \renewcommand{\arraystretch}{1.65}
  \begin{tabular}{l r r r r r r r r r r c r r r r r r r r r r}
    \toprule

    & \multicolumn{10}{c}{\textcolor{dpd2}{\textbf{2007Q2 -- 2019Q4}}}
    && \multicolumn{10}{c}{\textcolor{dpd2}{\textbf{2022Q1 -- 2025Q1}}}
    \\
    \cmidrule(lr){2-11}
    \cmidrule(lr){13-22}
    & FAAR & RF & LGB & LGB+ & LGB$^{\texttt{A}}+$ & KRR & NN & RR & SPF & TPFN 
    && FAAR & RF & LGB & LGB+ & LGB$^{\texttt{A}}+$ & KRR & NN & RR & SPF & TPFN \\

    \midrule
    \addlinespace[0.5em]
    \multicolumn{22}{l}{\textcolor{dpd2}{\textbf{Panel A: Squared Error}}} \\
    \addlinespace[0.3em]

      Return
      & -1.06 & {\textbf{0.13}} & 0.03 & -0.22 & -0.19 & -0.09 & -0.16 & 0.05 & {\color{ForestGreen}\textbf{0.13}} & 0.03 &  & -0.90 & -0.47 & -0.67 & -0.37 & -0.57 & -0.32 & -0.64 & {\color{ForestGreen}\textbf{0.13}} & -0.73 & {\textbf{-0.14}} \\
      \rowcolor{RowAlt}
      Sharpe
      & -0.80 & {\textbf{0.28}} & 0.04 & -0.31 & -0.25 & -0.25 & -0.40 & 0.22 & {\color{ForestGreen}\textbf{0.42}} & 0.11 &  & -1.06 & -0.95 & -0.86 & -0.76 & -0.79 & -0.39 & -0.60 & {\color{ForestGreen}\textbf{0.31}} & -0.84 & {\textbf{-0.23}} \\
      Sortino
      & -0.80 & {\textbf{0.76}} & 0.10 & -0.33 & -0.28 & -0.30 & -0.44 & 0.43 & {\color{ForestGreen}\textbf{1.46}} & 0.16 &  & -1.05 & -0.92 & -0.83 & -0.79 & -0.82 & -0.50 & -0.66 & {\color{ForestGreen}\textbf{0.62}} & -0.86 & {\textbf{-0.30}} \\
      \rowcolor{RowAlt}
      Omega
      & 0.23 & {\textbf{2.02}} & 1.11 & 0.43 & 0.56 & 0.65 & 0.47 & 1.51 & {\color{ForestGreen}\textbf{2.95}} & 1.19 &  & 0.21 & 0.20 & 0.20 & 0.30 & 0.32 & 0.63 & 0.43 & {\color{ForestGreen}\textbf{1.52}} & 0.29 & {\textbf{0.73}} \\
      MaxDD
      & -59.81 & {\color{ForestGreen}\textbf{-0.31}} & -9.86 & -15.15 & -16.08 & -6.95 & -10.71 & -0.57 & {\textbf{-0.55}} & -0.74 &  & -12.61 & -6.54 & -10.78 & {\textbf{-3.65}} & -9.63 & -9.93 & -13.15 & {\color{ForestGreen}\textbf{-0.71}} & -9.10 & -5.99 \\
      \rowcolor{RowAlt}
      Edge
      & {\textbf{0.90}} & 0.07 & {\color{ForestGreen}\textbf{1.30}} & 0.01 & 0.06 & 0.01 & 0.03 & 0.02 & 0.08 & 0.00 &  & {\color{ForestGreen}\textbf{0.35}} & 0.00 & 0.00 & 0.01 & 0.00 & {\textbf{0.28}} & 0.00 & 0.07 & 0.06 & 0.01 \\

    \addlinespace[0.5em]
    \midrule
    \addlinespace[0.5em]
    \multicolumn{22}{l}{\textcolor{dpd2}{\textbf{Panel B: Absolute Error}}} \\
    \addlinespace[0.3em]

      Return
      & -0.57 & {\textbf{0.04}} & -0.09 & -0.10 & -0.08 & -0.04 & -0.09 & 0.03 & {\color{ForestGreen}\textbf{0.09}} & 0.02 &  & -0.54 & -0.35 & -0.53 & -0.20 & -0.43 & -0.34 & -0.46 & {\color{ForestGreen}\textbf{-0.07}} & -0.36 & {\textbf{-0.20}} \\
      \rowcolor{RowAlt}
      Sharpe
      & -0.85 & {\textbf{0.19}} & -0.24 & -0.29 & -0.21 & -0.15 & -0.30 & 0.19 & {\color{ForestGreen}\textbf{0.52}} & 0.12 &  & -0.97 & -0.98 & -1.34 & -0.73 & -0.98 & -0.60 & -0.75 & {\color{ForestGreen}\textbf{-0.26}} & -0.71 & {\textbf{-0.49}} \\
      Sortino
      & -0.87 & 0.29 & -0.32 & -0.32 & -0.24 & -0.21 & -0.40 & {\textbf{0.30}} & {\color{ForestGreen}\textbf{0.86}} & 0.17 &  & -1.01 & -0.96 & -1.16 & -0.79 & -0.97 & -0.72 & -0.80 & {\color{ForestGreen}\textbf{-0.36}} & -0.79 & {\textbf{-0.59}} \\
      \rowcolor{RowAlt}
      Omega
      & 0.31 & {\textbf{1.31}} & 0.72 & 0.60 & 0.72 & 0.82 & 0.67 & 1.27 & {\color{ForestGreen}\textbf{1.91}} & 1.16 &  & 0.27 & 0.23 & 0.11 & 0.38 & 0.29 & 0.51 & 0.39 & {\color{ForestGreen}\textbf{0.74}} & 0.42 & {\textbf{0.56}} \\
      MaxDD
      & -33.45 & {\textbf{-0.68}} & -6.90 & -7.47 & -8.72 & -3.97 & -6.16 & -0.72 & {\color{ForestGreen}\textbf{-0.31}} & -0.93 &  & -8.25 & -5.03 & -7.74 & {\textbf{-3.18}} & -7.10 & -7.73 & -8.70 & {\color{ForestGreen}\textbf{-3.15}} & -5.94 & -5.18 \\
      \rowcolor{RowAlt}
      Edge
      & {\color{ForestGreen}\textbf{1.01}} & 0.09 & {\textbf{0.51}} & 0.09 & 0.24 & 0.07 & 0.19 & 0.03 & 0.25 & 0.01 &  & {\color{ForestGreen}\textbf{0.72}} & 0.00 & 0.00 & 0.10 & 0.00 & 0.19 & 0.00 & 0.02 & {\textbf{0.25}} & 0.15 \\

    \addlinespace[0.5em]
    \midrule
    \addlinespace[0.5em]
    \multicolumn{22}{l}{\textcolor{dpd2}{\textbf{Panel C: Classical Forecast Accuracy}}} \\
    \addlinespace[0.3em]

      RMSE
      & 1.43 & {\textbf{0.93}} & 0.98 & 1.10 & 1.09 & 1.04 & 1.08 & 0.97 & {\color{ForestGreen}\textbf{0.93}} & 0.99 &  & 1.38 & 1.21 & 1.29 & 1.17 & 1.25 & 1.15 & 1.28 & {\color{ForestGreen}\textbf{0.93}} & 1.31 & {\textbf{1.07}} \\
      \rowcolor{RowAlt}
      MAE
      & 1.57 & {\textbf{0.96}} & 1.09 & 1.10 & 1.08 & 1.04 & 1.09 & 0.97 & {\color{ForestGreen}\textbf{0.91}} & 0.98 &  & 1.54 & 1.35 & 1.53 & 1.20 & 1.43 & 1.34 & 1.46 & {\color{ForestGreen}\textbf{1.07}} & 1.36 & {\textbf{1.20}} \\
      $\rho(1)$
      & {\textbf{0.09}} & 0.32 & 0.30 & 0.14 & {\color{ForestGreen}\textbf{0.08}} & 0.49 & 0.45 & 0.47 & 0.42 & 0.41 &  & {\color{ForestGreen}\textbf{0.29}} & 0.58 & 0.50 & 0.47 & 0.52 & 0.56 & 0.57 & {\textbf{0.43}} & 0.57 & 0.48 \\
      \rowcolor{RowAlt}
      DM $t$-stat
      & -2.30 & {\textbf{1.00}} & 0.15 & -0.84 & -0.69 & -0.77 & -1.49 & 0.78 & {\color{ForestGreen}\textbf{1.65}} & 0.49 &  & -1.93 & -1.38 & -1.21 & -1.38 & -1.09 & -0.60 & -0.92 & {\color{ForestGreen}\textbf{0.45}} & -1.40 & {\textbf{-0.34}} \\

    \addlinespace[0.3em]
    \bottomrule
  \end{tabular}%
  }

  \vspace{0.25em}
  \parbox{\linewidth}{\scriptsize
    \textit{Notes}: Panels A--B report risk-adjusted metrics; Panel C reports classical forecast accuracy metrics. $\rho(1)$ = first-order autocorrelation of errors.
    Best: \textcolor{ForestGreen}{\textbf{bold green}}; second-best: \textbf{bold}.
  }
  }

\end{table}
\end{landscape}


\begin{landscape}
\begin{table}[t!]
  \centering
  \caption{\normalsize GDP ($h=4$)}
  \vspace*{-0.65em}
  \label{tab:gdp_h4}

  {\fontfamily{phv}\selectfont
  \resizebox{\linewidth}{!}{%
  \scriptsize
  \setlength{\tabcolsep}{0.3em}
  \renewcommand{\arraystretch}{1.65}
  \begin{tabular}{l r r r r r r r r r r c r r r r r r r r r r}
    \toprule

    & \multicolumn{10}{c}{\textcolor{dpd2}{\textbf{2007Q2 -- 2019Q4}}}
    && \multicolumn{10}{c}{\textcolor{dpd2}{\textbf{2022Q1 -- 2025Q1}}}
    \\
    \cmidrule(lr){2-11}
    \cmidrule(lr){13-22}
    & FAAR & RF & LGB & LGB+ & LGB$^{\texttt{A}}+$ & KRR & NN & RR & SPF & TPFN 
    && FAAR & RF & LGB & LGB+ & LGB$^{\texttt{A}}+$ & KRR & NN & RR & SPF & TPFN \\

    \midrule
    \addlinespace[0.5em]
    \multicolumn{22}{l}{\textcolor{dpd2}{\textbf{Panel A: Squared Error}}} \\
    \addlinespace[0.3em]

      Return
      & -0.45 & 0.01 & -0.04 & -0.09 & -0.29 & 0.11 & 0.10 & 0.05 & {\textbf{0.14}} & {\color{ForestGreen}\textbf{0.16}} &  & -0.06 & 0.57 & 0.23 & 0.62 & {\textbf{0.69}} & 0.56 & 0.64 & 0.57 & 0.43 & {\color{ForestGreen}\textbf{0.71}} \\
      \rowcolor{RowAlt}
      Sharpe
      & -0.75 & 0.04 & -0.15 & -0.50 & -0.80 & {\textbf{0.46}} & 0.21 & 0.24 & {\color{ForestGreen}\textbf{0.64}} & 0.44 &  & -0.53 & 0.46 & 0.17 & {\textbf{0.57}} & 0.50 & 0.49 & 0.46 & 0.49 & 0.46 & {\color{ForestGreen}\textbf{0.59}} \\
      Sortino
      & -0.72 & 0.06 & -0.21 & -0.54 & -0.78 & {\textbf{1.58}} & 0.44 & 0.51 & {\color{ForestGreen}\textbf{1.97}} & 1.28 &  & -0.58 & 2.67 & 0.32 & {\textbf{8.50}} & 4.50 & 4.30 & 2.47 & 4.84 & 2.25 & {\color{ForestGreen}\textbf{9.43}} \\
      \rowcolor{RowAlt}
      Omega
      & 0.16 & 1.07 & 0.78 & 0.43 & 0.25 & {\textbf{2.65}} & 1.56 & 1.60 & {\color{ForestGreen}\textbf{3.65}} & 2.51 &  & 0.38 & 3.21 & 1.43 & {\textbf{8.55}} & 4.23 & 4.53 & 3.29 & 5.16 & 3.15 & {\color{ForestGreen}\textbf{9.09}} \\
      MaxDD
      & -24.38 & -0.91 & -5.64 & -5.15 & -14.63 & {\color{ForestGreen}\textbf{-0.12}} & -0.48 & -0.51 & {\textbf{-0.19}} & -0.21 &  & -0.68 & -0.29 & -0.71 & {\textbf{-0.10}} & -0.21 & -0.20 & -0.28 & -0.17 & -0.23 & {\color{ForestGreen}\textbf{-0.07}} \\
      \rowcolor{RowAlt}
      Edge
      & 0.04 & 0.00 & 0.30 & 0.04 & 0.09 & 0.01 & {\color{ForestGreen}\textbf{1.18}} & 0.00 & {\textbf{1.10}} & 0.55 &  & 0.02 & 0.13 & 0.00 & 0.03 & {\textbf{0.20}} & 0.00 & 0.00 & 0.00 & {\color{ForestGreen}\textbf{0.23}} & 0.00 \\

    \addlinespace[0.5em]
    \midrule
    \addlinespace[0.5em]
    \multicolumn{22}{l}{\textcolor{dpd2}{\textbf{Panel B: Absolute Error}}} \\
    \addlinespace[0.3em]

      Return
      & -0.33 & 0.03 & -0.05 & -0.03 & -0.16 & {\textbf{0.07}} & -0.02 & 0.02 & {\color{ForestGreen}\textbf{0.10}} & 0.06 &  & -0.05 & 0.06 & -0.09 & {\textbf{0.20}} & 0.16 & 0.02 & 0.12 & 0.04 & 0.09 & {\color{ForestGreen}\textbf{0.21}} \\
      \rowcolor{RowAlt}
      Sharpe
      & -0.97 & 0.14 & -0.19 & -0.18 & -0.57 & {\textbf{0.42}} & -0.06 & 0.15 & {\color{ForestGreen}\textbf{0.72}} & 0.25 &  & -0.34 & 0.09 & -0.13 & {\color{ForestGreen}\textbf{0.43}} & 0.20 & 0.04 & 0.15 & 0.07 & 0.20 & {\textbf{0.38}} \\
      Sortino
      & -0.91 & 0.21 & -0.24 & -0.24 & -0.64 & {\textbf{0.71}} & -0.08 & 0.22 & {\color{ForestGreen}\textbf{1.41}} & 0.39 &  & -0.41 & 0.17 & -0.19 & {\color{ForestGreen}\textbf{1.31}} & 0.54 & 0.09 & 0.34 & 0.17 & 0.34 & {\textbf{1.22}} \\
      \rowcolor{RowAlt}
      Omega
      & 0.20 & 1.20 & 0.78 & 0.79 & 0.47 & {\textbf{1.70}} & 0.92 & 1.21 & {\color{ForestGreen}\textbf{2.46}} & 1.40 &  & 0.63 & 1.13 & 0.83 & {\color{ForestGreen}\textbf{2.14}} & 1.38 & 1.06 & 1.26 & 1.12 & 1.29 & {\textbf{1.97}} \\
      MaxDD
      & -17.34 & -0.75 & -5.11 & -2.65 & -7.84 & {\textbf{-0.29}} & -0.98 & -0.70 & {\color{ForestGreen}\textbf{-0.17}} & -0.46 &  & -1.20 & -4.59 & -5.82 & {\color{ForestGreen}\textbf{-0.42}} & -0.78 & -0.92 & -0.92 & -0.78 & -0.68 & {\textbf{-0.42}} \\
      \rowcolor{RowAlt}
      Edge
      & 0.14 & 0.06 & 0.39 & 0.23 & 0.30 & 0.02 & 0.28 & 0.01 & {\color{ForestGreen}\textbf{1.23}} & {\textbf{0.47}} &  & 0.24 & 0.21 & 0.00 & 0.12 & {\textbf{0.31}} & 0.00 & 0.00 & 0.00 & {\color{ForestGreen}\textbf{0.79}} & 0.00 \\

    \addlinespace[0.5em]
    \midrule
    \addlinespace[0.5em]
    \multicolumn{22}{l}{\textcolor{dpd2}{\textbf{Panel C: Classical Forecast Accuracy}}} \\
    \addlinespace[0.3em]

      RMSE
      & 1.20 & 1.00 & 1.02 & 1.04 & 1.14 & 0.94 & 0.95 & 0.97 & {\textbf{0.93}} & {\color{ForestGreen}\textbf{0.92}} &  & 1.03 & 0.65 & 0.88 & 0.61 & {\textbf{0.56}} & 0.67 & 0.60 & 0.66 & 0.76 & {\color{ForestGreen}\textbf{0.54}} \\
      \rowcolor{RowAlt}
      MAE
      & 1.33 & 0.97 & 1.05 & 1.03 & 1.16 & {\textbf{0.93}} & 1.02 & 0.98 & {\color{ForestGreen}\textbf{0.90}} & 0.94 &  & 1.05 & 0.94 & 1.09 & {\textbf{0.80}} & 0.84 & 0.98 & 0.88 & 0.96 & 0.91 & {\color{ForestGreen}\textbf{0.79}} \\
      $\rho(1)$
      & 0.57 & 0.53 & 0.52 & 0.58 & 0.59 & 0.52 & {\textbf{0.50}} & 0.55 & 0.53 & {\color{ForestGreen}\textbf{0.36}} &  & {\color{ForestGreen}\textbf{-0.01}} & 0.31 & 0.29 & {\textbf{0.20}} & 0.44 & 0.41 & 0.47 & 0.42 & 0.34 & 0.25 \\
      \rowcolor{RowAlt}
      DM $t$-stat
      & -2.02 & 0.16 & -0.55 & -1.90 & -1.99 & {\textbf{1.63}} & 0.54 & 0.90 & {\color{ForestGreen}\textbf{2.57}} & 1.22 &  & -0.97 & 0.83 & 0.30 & {\textbf{1.03}} & 0.90 & 0.89 & 0.82 & 0.88 & 0.85 & {\color{ForestGreen}\textbf{1.06}} \\

    \addlinespace[0.3em]
    \bottomrule
  \end{tabular}%
  }

  \vspace{0.25em}
  \parbox{\linewidth}{\scriptsize
    \textit{Notes}: Panels A--B report risk-adjusted metrics; Panel C reports classical forecast accuracy metrics. $\rho(1)$ = first-order autocorrelation of errors.
    Best: \textcolor{ForestGreen}{\textbf{bold green}}; second-best: \textbf{bold}.
  }
  }

\end{table}
\end{landscape}


\begin{landscape}
\begin{table}[t!]
  \centering
  \caption{\normalsize Unemployment Rate ($h=1$)}
  \vspace*{-0.65em}
  \label{tab:ur_h1}

  {\fontfamily{phv}\selectfont
  \resizebox{\linewidth}{!}{%
  \scriptsize
  \setlength{\tabcolsep}{0.3em}
  \renewcommand{\arraystretch}{1.65}
  \begin{tabular}{l r r r r r r r r r r c r r r r r r r r r r}
    \toprule

    & \multicolumn{10}{c}{\textcolor{dpd2}{\textbf{2007Q2 -- 2019Q4}}}
    && \multicolumn{10}{c}{\textcolor{dpd2}{\textbf{2022Q1 -- 2025Q1}}}
    \\
    \cmidrule(lr){2-11}
    \cmidrule(lr){13-22}
    & FAAR & RF & LGB & LGB+ & LGB$^{\texttt{A}}+$ & KRR & NN & RR & SPF & TPFN 
    && FAAR & RF & LGB & LGB+ & LGB$^{\texttt{A}}+$ & KRR & NN & RR & SPF & TPFN \\

    \midrule
    \addlinespace[0.5em]
    \multicolumn{22}{l}{\textcolor{dpd2}{\textbf{Panel A: Squared Error}}} \\
    \addlinespace[0.3em]

      Return
      & 0.21 & 0.27 & 0.26 & {\color{ForestGreen}\textbf{0.55}} & {\textbf{0.54}} & -0.67 & 0.11 & -0.07 & -0.14 & 0.29 &  & -2.77 & -1.05 & -1.27 & -0.94 & -0.99 & -8.64 & -1.00 & {\textbf{-0.30}} & -0.31 & {\color{ForestGreen}\textbf{0.18}} \\
      \rowcolor{RowAlt}
      Sharpe
      & 0.23 & {\color{ForestGreen}\textbf{0.80}} & 0.57 & 0.69 & 0.68 & -0.38 & 0.20 & -0.17 & -0.19 & {\textbf{0.70}} &  & -0.75 & -0.66 & -0.54 & -0.56 & -0.79 & -1.22 & -0.56 & {\textbf{-0.33}} & -0.33 & {\color{ForestGreen}\textbf{0.20}} \\
      Sortino
      & 0.47 & 2.84 & 1.48 & {\color{ForestGreen}\textbf{7.73}} & {\textbf{6.38}} & -0.43 & 0.33 & -0.21 & -0.20 & 1.89 &  & -0.73 & -0.68 & -0.55 & -0.59 & -0.83 & -1.07 & -0.58 & {\textbf{-0.38}} & -0.40 & {\color{ForestGreen}\textbf{0.26}} \\
      \rowcolor{RowAlt}
      Omega
      & 1.70 & 4.80 & 3.04 & {\color{ForestGreen}\textbf{11.48}} & {\textbf{10.50}} & 0.43 & 1.46 & 0.76 & 0.66 & 3.81 &  & 0.20 & 0.29 & 0.29 & 0.38 & 0.33 & 0.02 & 0.37 & {\textbf{0.61}} & 0.60 & {\color{ForestGreen}\textbf{1.34}} \\
      MaxDD
      & -0.36 & -0.09 & -0.31 & {\color{ForestGreen}\textbf{-0.03}} & {\textbf{-0.05}} & -37.30 & -0.45 & -7.26 & -16.69 & -0.18 &  & -44.83 & -15.18 & -18.74 & -18.92 & -16.96 & -65.90 & -17.79 & {\textbf{-8.29}} & -9.04 & {\color{ForestGreen}\textbf{-4.65}} \\
      \rowcolor{RowAlt}
      Edge
      & 0.09 & 0.03 & 0.12 & {\color{ForestGreen}\textbf{0.55}} & 0.05 & {\textbf{0.27}} & 0.08 & 0.03 & 0.23 & 0.17 &  & {\textbf{0.04}} & 0.01 & 0.03 & 0.00 & 0.02 & 0.00 & 0.04 & 0.00 & 0.03 & {\color{ForestGreen}\textbf{0.23}} \\

    \addlinespace[0.5em]
    \midrule
    \addlinespace[0.5em]
    \multicolumn{22}{l}{\textcolor{dpd2}{\textbf{Panel B: Absolute Error}}} \\
    \addlinespace[0.3em]

      Return
      & 0.07 & 0.26 & 0.21 & {\textbf{0.30}} & {\color{ForestGreen}\textbf{0.31}} & -0.23 & 0.12 & 0.02 & 0.05 & 0.24 &  & -0.52 & -0.32 & -0.28 & -0.30 & -0.53 & -1.87 & -0.41 & -0.17 & {\textbf{-0.09}} & {\color{ForestGreen}\textbf{0.07}} \\
      \rowcolor{RowAlt}
      Sharpe
      & 0.20 & {\textbf{1.12}} & 0.72 & 1.06 & {\color{ForestGreen}\textbf{1.14}} & -0.38 & 0.37 & 0.06 & 0.13 & 0.90 &  & -0.47 & -0.46 & -0.33 & -0.42 & -0.88 & -1.48 & -0.63 & -0.37 & {\textbf{-0.15}} & {\color{ForestGreen}\textbf{0.14}} \\
      Sortino
      & 0.30 & 2.94 & 1.42 & {\color{ForestGreen}\textbf{3.71}} & {\textbf{3.42}} & -0.44 & 0.62 & 0.08 & 0.17 & 2.01 &  & -0.50 & -0.53 & -0.37 & -0.48 & -0.93 & -1.23 & -0.69 & -0.46 & {\textbf{-0.20}} & {\color{ForestGreen}\textbf{0.20}} \\
      \rowcolor{RowAlt}
      Omega
      & 1.34 & 4.46 & 2.63 & {\color{ForestGreen}\textbf{5.15}} & {\textbf{5.00}} & 0.56 & 1.71 & 1.08 & 1.22 & 3.42 &  & 0.48 & 0.51 & 0.59 & 0.56 & 0.36 & 0.07 & 0.42 & 0.61 & {\textbf{0.82}} & {\color{ForestGreen}\textbf{1.20}} \\
      MaxDD
      & -0.52 & {\textbf{-0.11}} & -0.49 & -0.12 & {\color{ForestGreen}\textbf{-0.11}} & -13.22 & -0.56 & -3.09 & -5.22 & -0.17 &  & -12.65 & -5.92 & -5.11 & -7.86 & -9.03 & -17.89 & -7.03 & {\textbf{-4.41}} & -4.90 & {\color{ForestGreen}\textbf{-1.00}} \\
      \rowcolor{RowAlt}
      Edge
      & 0.11 & 0.16 & 0.35 & 0.35 & 0.05 & {\textbf{0.40}} & 0.21 & 0.11 & {\color{ForestGreen}\textbf{0.48}} & 0.37 &  & {\textbf{0.29}} & 0.10 & 0.20 & 0.02 & 0.12 & 0.00 & 0.07 & 0.00 & {\color{ForestGreen}\textbf{0.30}} & 0.17 \\

    \addlinespace[0.5em]
    \midrule
    \addlinespace[0.5em]
    \multicolumn{22}{l}{\textcolor{dpd2}{\textbf{Panel C: Classical Forecast Accuracy}}} \\
    \addlinespace[0.3em]

      RMSE
      & 0.89 & 0.86 & 0.86 & {\color{ForestGreen}\textbf{0.67}} & {\textbf{0.68}} & 1.29 & 0.94 & 1.04 & 1.07 & 0.84 &  & 1.94 & 1.43 & 1.51 & 1.39 & 1.41 & 3.11 & 1.41 & {\textbf{1.14}} & 1.15 & {\color{ForestGreen}\textbf{0.91}} \\
      \rowcolor{RowAlt}
      MAE
      & 0.93 & 0.74 & 0.79 & {\textbf{0.70}} & {\color{ForestGreen}\textbf{0.69}} & 1.23 & 0.88 & 0.98 & 0.95 & 0.76 &  & 1.52 & 1.32 & 1.28 & 1.30 & 1.53 & 2.87 & 1.41 & 1.17 & {\textbf{1.09}} & {\color{ForestGreen}\textbf{0.93}} \\
      $\rho(1)$
      & 0.06 & 0.27 & {\textbf{0.02}} & 0.04 & 0.08 & -0.12 & {\color{ForestGreen}\textbf{-0.01}} & 0.52 & 0.60 & 0.21 &  & 0.72 & 0.29 & {\textbf{0.09}} & 0.56 & 0.66 & 0.39 & 0.21 & 0.43 & 0.40 & {\color{ForestGreen}\textbf{0.08}} \\
      \rowcolor{RowAlt}
      DM $t$-stat
      & 0.82 & {\color{ForestGreen}\textbf{2.53}} & 1.73 & 1.86 & {\textbf{1.91}} & -1.18 & 0.55 & -0.55 & -0.51 & 1.88 &  & -1.04 & -1.05 & -1.00 & -0.80 & -1.51 & -2.09 & -0.88 & {\textbf{-0.55}} & -0.60 & {\color{ForestGreen}\textbf{0.42}} \\

    \addlinespace[0.3em]
    \bottomrule
  \end{tabular}%
  }

  \vspace{0.25em}
  \parbox{\linewidth}{\scriptsize
    \textit{Notes}: Panels A--B report risk-adjusted metrics; Panel C reports classical forecast accuracy metrics. $\rho(1)$ = first-order autocorrelation of errors.
    Best: \textcolor{ForestGreen}{\textbf{bold green}}; second-best: \textbf{bold}.
  }
  }

\end{table}
\end{landscape}


\begin{landscape}
\begin{table}[t!]
  \centering
  \caption{\normalsize Unemployment Rate ($h=2$)}
  \vspace*{-0.65em}
  \label{tab:ur_h2}

  {\fontfamily{phv}\selectfont
  \resizebox{\linewidth}{!}{%
  \scriptsize
  \setlength{\tabcolsep}{0.3em}
  \renewcommand{\arraystretch}{1.65}
  \begin{tabular}{l r r r r r r r r r r c r r r r r r r r r r}
    \toprule

    & \multicolumn{10}{c}{\textcolor{dpd2}{\textbf{2007Q2 -- 2019Q4}}}
    && \multicolumn{10}{c}{\textcolor{dpd2}{\textbf{2022Q1 -- 2025Q1}}}
    \\
    \cmidrule(lr){2-11}
    \cmidrule(lr){13-22}
    & FAAR & RF & LGB & LGB+ & LGB$^{\texttt{A}}+$ & KRR & NN & RR & SPF & TPFN 
    && FAAR & RF & LGB & LGB+ & LGB$^{\texttt{A}}+$ & KRR & NN & RR & SPF & TPFN \\

    \midrule
    \addlinespace[0.5em]
    \multicolumn{22}{l}{\textcolor{dpd2}{\textbf{Panel A: Squared Error}}} \\
    \addlinespace[0.3em]

      Return
      & 0.12 & 0.09 & 0.15 & {\color{ForestGreen}\textbf{0.50}} & {\textbf{0.29}} & -0.74 & 0.19 & 0.02 & 0.13 & 0.25 &  & -2.87 & -2.56 & -7.49 & -1.86 & -0.73 & -5.56 & -0.67 & {\textbf{-0.04}} & {\color{ForestGreen}\textbf{0.16}} & -0.31 \\
      \rowcolor{RowAlt}
      Sharpe
      & 0.25 & 0.29 & 0.34 & {\color{ForestGreen}\textbf{0.82}} & {\textbf{0.68}} & -0.44 & 0.24 & 0.13 & 0.64 & 0.58 &  & -1.40 & -0.77 & -0.83 & -0.65 & -0.46 & -1.34 & -0.46 & {\textbf{-0.06}} & {\color{ForestGreen}\textbf{0.20}} & -0.37 \\
      Sortino
      & 0.47 & 0.52 & 0.65 & {\color{ForestGreen}\textbf{10.98}} & {\textbf{5.06}} & -0.51 & 0.61 & 0.22 & 1.73 & 1.71 &  & -1.19 & -0.75 & -0.79 & -0.65 & -0.49 & -1.14 & -0.51 & {\textbf{-0.07}} & {\color{ForestGreen}\textbf{0.29}} & -0.46 \\
      \rowcolor{RowAlt}
      Omega
      & 1.56 & 1.67 & 1.80 & {\color{ForestGreen}\textbf{14.31}} & {\textbf{6.82}} & 0.39 & 1.70 & 1.28 & 3.09 & 3.67 &  & 0.10 & 0.16 & 0.08 & 0.24 & 0.45 & 0.01 & 0.52 & {\textbf{0.93}} & {\color{ForestGreen}\textbf{1.31}} & 0.64 \\
      MaxDD
      & -0.43 & -0.70 & -0.32 & {\color{ForestGreen}\textbf{-0.02}} & {\textbf{-0.09}} & -57.01 & -0.50 & -0.67 & -0.54 & -0.21 &  & -41.63 & -35.13 & -101.22 & -28.98 & -14.60 & -56.13 & -16.05 & {\textbf{-4.82}} & {\color{ForestGreen}\textbf{-0.90}} & -8.01 \\
      \rowcolor{RowAlt}
      Edge
      & 0.12 & 0.01 & 0.03 & {\textbf{0.32}} & 0.00 & {\color{ForestGreen}\textbf{0.56}} & 0.02 & 0.01 & 0.09 & 0.00 &  & 0.02 & 0.00 & 0.01 & {\textbf{0.03}} & 0.01 & 0.00 & {\color{ForestGreen}\textbf{0.06}} & 0.03 & 0.00 & 0.00 \\

    \addlinespace[0.5em]
    \midrule
    \addlinespace[0.5em]
    \multicolumn{22}{l}{\textcolor{dpd2}{\textbf{Panel B: Absolute Error}}} \\
    \addlinespace[0.3em]

      Return
      & -0.02 & 0.10 & 0.09 & {\color{ForestGreen}\textbf{0.31}} & 0.16 & -0.42 & 0.02 & 0.01 & 0.10 & {\textbf{0.17}} &  & -0.98 & -0.63 & -1.17 & -0.49 & -0.21 & -1.60 & -0.20 & {\textbf{0.03}} & {\color{ForestGreen}\textbf{0.11}} & -0.21 \\
      \rowcolor{RowAlt}
      Sharpe
      & -0.05 & 0.41 & 0.29 & {\color{ForestGreen}\textbf{1.07}} & {\textbf{0.69}} & -0.58 & 0.06 & 0.06 & 0.55 & 0.69 &  & -1.38 & -0.67 & -0.74 & -0.55 & -0.31 & -1.89 & -0.28 & {\textbf{0.07}} & {\color{ForestGreen}\textbf{0.20}} & -0.40 \\
      Sortino
      & -0.06 & 0.60 & 0.43 & {\color{ForestGreen}\textbf{3.15}} & {\textbf{1.44}} & -0.63 & 0.09 & 0.08 & 1.08 & 1.24 &  & -1.20 & -0.70 & -0.74 & -0.60 & -0.37 & -1.40 & -0.35 & {\textbf{0.10}} & {\color{ForestGreen}\textbf{0.30}} & -0.52 \\
      \rowcolor{RowAlt}
      Omega
      & 0.94 & 1.75 & 1.45 & {\color{ForestGreen}\textbf{4.67}} & {\textbf{2.62}} & 0.40 & 1.10 & 1.09 & 2.03 & 2.52 &  & 0.18 & 0.35 & 0.27 & 0.43 & 0.64 & 0.02 & 0.72 & {\textbf{1.10}} & {\color{ForestGreen}\textbf{1.30}} & 0.61 \\
      MaxDD
      & -6.34 & -0.88 & -0.90 & {\color{ForestGreen}\textbf{-0.11}} & {\textbf{-0.42}} & -28.56 & -0.94 & -2.97 & -0.50 & -0.49 &  & -15.68 & -10.05 & -18.33 & -9.47 & -5.82 & -17.89 & -7.66 & {\textbf{-3.32}} & {\color{ForestGreen}\textbf{-0.91}} & -5.21 \\
      \rowcolor{RowAlt}
      Edge
      & 0.31 & 0.09 & 0.13 & {\textbf{0.47}} & 0.04 & {\color{ForestGreen}\textbf{0.47}} & 0.08 & 0.07 & 0.19 & 0.02 &  & 0.10 & 0.00 & 0.08 & 0.07 & 0.10 & 0.00 & {\color{ForestGreen}\textbf{0.33}} & {\textbf{0.11}} & 0.08 & 0.00 \\

    \addlinespace[0.5em]
    \midrule
    \addlinespace[0.5em]
    \multicolumn{22}{l}{\textcolor{dpd2}{\textbf{Panel C: Classical Forecast Accuracy}}} \\
    \addlinespace[0.3em]

      RMSE
      & 0.94 & 0.95 & 0.92 & {\color{ForestGreen}\textbf{0.71}} & {\textbf{0.84}} & 1.32 & 0.90 & 0.99 & 0.94 & 0.87 &  & 1.97 & 1.89 & 2.91 & 1.69 & 1.31 & 2.56 & 1.29 & {\textbf{1.02}} & {\color{ForestGreen}\textbf{0.92}} & 1.14 \\
      \rowcolor{RowAlt}
      MAE
      & 1.02 & 0.90 & 0.91 & {\color{ForestGreen}\textbf{0.69}} & 0.84 & 1.42 & 0.98 & 0.99 & 0.90 & {\textbf{0.83}} &  & 1.98 & 1.63 & 2.17 & 1.49 & 1.21 & 2.60 & 1.20 & {\textbf{0.97}} & {\color{ForestGreen}\textbf{0.89}} & 1.21 \\
      $\rho(1)$
      & 0.47 & 0.66 & 0.50 & 0.38 & 0.48 & {\color{ForestGreen}\textbf{-0.11}} & {\textbf{0.20}} & 0.69 & 0.63 & 0.57 &  & {\textbf{0.42}} & 0.62 & 0.67 & 0.65 & 0.68 & 0.52 & 0.52 & 0.46 & {\color{ForestGreen}\textbf{0.40}} & 0.71 \\
      \rowcolor{RowAlt}
      DM $t$-stat
      & 0.63 & 0.89 & 1.21 & 1.76 & 1.77 & -1.28 & 0.59 & 0.39 & {\color{ForestGreen}\textbf{2.16}} & {\textbf{2.13}} &  & -2.00 & -1.15 & -1.21 & -1.03 & -0.70 & -2.45 & -0.72 & {\textbf{-0.10}} & {\color{ForestGreen}\textbf{0.33}} & -0.61 \\

    \addlinespace[0.3em]
    \bottomrule
  \end{tabular}%
  }

  \vspace{0.25em}
  \parbox{\linewidth}{\scriptsize
    \textit{Notes}: Panels A--B report risk-adjusted metrics; Panel C reports classical forecast accuracy metrics. $\rho(1)$ = first-order autocorrelation of errors.
    Best: \textcolor{ForestGreen}{\textbf{bold green}}; second-best: \textbf{bold}.
  }
  }

\end{table}
\end{landscape}


\begin{landscape}
\begin{table}[t!]
  \centering
  \caption{\normalsize Unemployment Rate ($h=4$)}
  \vspace*{-0.65em}
  \label{tab:ur_h4}

  {\fontfamily{phv}\selectfont
  \resizebox{\linewidth}{!}{%
  \scriptsize
  \setlength{\tabcolsep}{0.3em}
  \renewcommand{\arraystretch}{1.65}
  \begin{tabular}{l r r r r r r r r r r c r r r r r r r r r r}
    \toprule

    & \multicolumn{10}{c}{\textcolor{dpd2}{\textbf{2007Q2 -- 2019Q4}}}
    && \multicolumn{10}{c}{\textcolor{dpd2}{\textbf{2022Q1 -- 2025Q1}}}
    \\
    \cmidrule(lr){2-11}
    \cmidrule(lr){13-22}
    & FAAR & RF & LGB & LGB+ & LGB$^{\texttt{A}}+$ & KRR & NN & RR & SPF & TPFN 
    && FAAR & RF & LGB & LGB+ & LGB$^{\texttt{A}}+$ & KRR & NN & RR & SPF & TPFN \\

    \midrule
    \addlinespace[0.5em]
    \multicolumn{22}{l}{\textcolor{dpd2}{\textbf{Panel A: Squared Error}}} \\
    \addlinespace[0.3em]

      Return
      & -0.15 & 0.20 & {\textbf{0.28}} & 0.25 & 0.01 & -0.69 & 0.18 & 0.11 & 0.16 & {\color{ForestGreen}\textbf{0.48}} &  & -2.90 & -1.11 & -1.15 & -1.42 & -0.36 & -3.01 & 0.33 & {\textbf{0.65}} & {\color{ForestGreen}\textbf{0.76}} & 0.40 \\
      \rowcolor{RowAlt}
      Sharpe
      & -0.36 & {\color{ForestGreen}\textbf{1.10}} & 0.67 & 0.74 & 0.05 & -0.47 & 0.25 & 0.47 & {\textbf{0.92}} & 0.72 &  & -0.89 & -0.59 & -0.62 & -0.75 & -0.32 & -1.43 & 0.32 & {\color{ForestGreen}\textbf{0.91}} & {\textbf{0.88}} & 0.51 \\
      Sortino
      & -0.39 & {\textbf{4.82}} & 1.86 & 1.98 & 0.09 & -0.56 & 0.51 & 1.03 & 2.16 & {\color{ForestGreen}\textbf{6.14}} &  & -0.84 & -0.63 & -0.66 & -0.76 & -0.38 & -1.20 & 0.63 & {\textbf{6.01}} & {\color{ForestGreen}\textbf{6.50}} & 1.36 \\
      \rowcolor{RowAlt}
      Omega
      & 0.52 & {\textbf{8.46}} & 4.69 & 4.97 & 1.09 & 0.41 & 1.79 & 2.71 & 4.44 & {\color{ForestGreen}\textbf{10.20}} &  & 0.01 & 0.35 & 0.35 & 0.27 & 0.57 & 0.04 & 1.59 & {\color{ForestGreen}\textbf{7.70}} & {\textbf{6.87}} & 2.34 \\
      MaxDD
      & -10.33 & {\color{ForestGreen}\textbf{-0.09}} & -1.65 & -0.24 & -0.78 & -53.00 & -0.50 & -0.32 & -0.47 & {\textbf{-0.12}} &  & -34.72 & -21.34 & -21.79 & -24.45 & -9.81 & -25.90 & -0.46 & {\textbf{-0.10}} & {\color{ForestGreen}\textbf{-0.06}} & -0.31 \\
      \rowcolor{RowAlt}
      Edge
      & 0.07 & 0.01 & 0.06 & 0.01 & 0.00 & {\textbf{0.75}} & 0.01 & 0.00 & 0.08 & {\color{ForestGreen}\textbf{4.45}} &  & 0.01 & 0.00 & 0.00 & 0.00 & 0.00 & 0.00 & {\textbf{0.10}} & {\color{ForestGreen}\textbf{0.22}} & 0.08 & 0.05 \\

    \addlinespace[0.5em]
    \midrule
    \addlinespace[0.5em]
    \multicolumn{22}{l}{\textcolor{dpd2}{\textbf{Panel B: Absolute Error}}} \\
    \addlinespace[0.3em]

      Return
      & -0.13 & 0.19 & 0.19 & {\textbf{0.22}} & -0.02 & -0.44 & 0.10 & 0.07 & 0.18 & {\color{ForestGreen}\textbf{0.35}} &  & -0.89 & -0.45 & -0.47 & -0.61 & -0.27 & -1.13 & 0.15 & {\textbf{0.31}} & {\color{ForestGreen}\textbf{0.41}} & 0.23 \\
      \rowcolor{RowAlt}
      Sharpe
      & -0.39 & {\color{ForestGreen}\textbf{1.21}} & 0.89 & {\textbf{1.02}} & -0.12 & -0.70 & 0.33 & 0.50 & 0.98 & 0.98 &  & -1.44 & -0.53 & -0.57 & -0.74 & -0.44 & -1.68 & 0.22 & {\color{ForestGreen}\textbf{0.81}} & {\textbf{0.74}} & 0.41 \\
      Sortino
      & -0.45 & {\color{ForestGreen}\textbf{3.21}} & 1.59 & 2.45 & -0.16 & -0.76 & 0.53 & 0.78 & 2.14 & {\textbf{2.98}} &  & -1.21 & -0.59 & -0.62 & -0.76 & -0.49 & -1.32 & 0.36 & {\textbf{2.12}} & {\color{ForestGreen}\textbf{2.47}} & 0.82 \\
      \rowcolor{RowAlt}
      Omega
      & 0.58 & {\color{ForestGreen}\textbf{4.99}} & 3.24 & 4.10 & 0.86 & 0.41 & 1.50 & 1.93 & 3.80 & {\textbf{4.84}} &  & 0.06 & 0.45 & 0.44 & 0.33 & 0.49 & 0.08 & 1.33 & {\textbf{3.02}} & {\color{ForestGreen}\textbf{3.19}} & 1.72 \\
      MaxDD
      & -8.34 & {\color{ForestGreen}\textbf{-0.14}} & -0.79 & -0.31 & -4.78 & -29.29 & -0.54 & -0.48 & -0.63 & {\textbf{-0.26}} &  & -10.75 & -9.74 & -9.72 & -10.92 & -6.03 & -11.81 & -0.52 & {\textbf{-0.27}} & {\color{ForestGreen}\textbf{-0.15}} & -0.38 \\
      \rowcolor{RowAlt}
      Edge
      & 0.26 & 0.11 & 0.19 & 0.07 & 0.02 & {\textbf{0.58}} & 0.08 & 0.03 & 0.18 & {\color{ForestGreen}\textbf{2.95}} &  & 0.08 & 0.00 & 0.02 & 0.02 & 0.00 & 0.00 & {\textbf{0.38}} & {\color{ForestGreen}\textbf{0.50}} & 0.35 & 0.34 \\

    \addlinespace[0.5em]
    \midrule
    \addlinespace[0.5em]
    \multicolumn{22}{l}{\textcolor{dpd2}{\textbf{Panel C: Classical Forecast Accuracy}}} \\
    \addlinespace[0.3em]

      RMSE
      & 1.07 & 0.90 & {\textbf{0.85}} & 0.87 & 0.99 & 1.30 & 0.91 & 0.95 & 0.91 & {\color{ForestGreen}\textbf{0.72}} &  & 1.98 & 1.45 & 1.46 & 1.56 & 1.16 & 2.00 & 0.82 & {\textbf{0.59}} & {\color{ForestGreen}\textbf{0.49}} & 0.77 \\
      \rowcolor{RowAlt}
      MAE
      & 1.13 & 0.81 & 0.81 & {\textbf{0.78}} & 1.02 & 1.44 & 0.90 & 0.93 & 0.82 & {\color{ForestGreen}\textbf{0.65}} &  & 1.89 & 1.45 & 1.47 & 1.61 & 1.27 & 2.13 & 0.85 & {\textbf{0.69}} & {\color{ForestGreen}\textbf{0.59}} & 0.77 \\
      $\rho(1)$
      & 0.65 & 0.74 & 0.66 & 0.73 & 0.76 & {\textbf{0.59}} & 0.60 & 0.76 & 0.73 & {\color{ForestGreen}\textbf{0.41}} &  & {\textbf{-0.06}} & 1.08 & 0.93 & 0.94 & 0.90 & {\color{ForestGreen}\textbf{0.02}} & 0.39 & 0.35 & 0.20 & 0.32 \\
      \rowcolor{RowAlt}
      DM $t$-stat
      & -1.26 & {\textbf{2.83}} & 1.60 & 1.94 & 0.11 & -1.61 & 0.91 & 1.50 & {\color{ForestGreen}\textbf{3.58}} & 1.72 &  & -1.60 & -0.97 & -2.26 & -2.93 & -0.43 & -2.24 & 0.45 & {\textbf{1.26}} & {\color{ForestGreen}\textbf{1.28}} & 0.87 \\

    \addlinespace[0.3em]
    \bottomrule
  \end{tabular}%
  }

  \vspace{0.25em}
  \parbox{\linewidth}{\scriptsize
    \textit{Notes}: Panels A--B report risk-adjusted metrics; Panel C reports classical forecast accuracy metrics. $\rho(1)$ = first-order autocorrelation of errors.
    Best: \textcolor{ForestGreen}{\textbf{bold green}}; second-best: \textbf{bold}.
  }
  }

\end{table}
\end{landscape}


\begin{landscape}
\begin{table}[t!]
  \centering
  \caption{\normalsize Inflation ($h=1$)}
  \vspace*{-0.65em}
  \label{tab:infl_h1}

  {\fontfamily{phv}\selectfont
  \resizebox{\linewidth}{!}{%
  \scriptsize
  \setlength{\tabcolsep}{0.3em}
  \renewcommand{\arraystretch}{1.65}
  \begin{tabular}{l r r r r r r r r r r r c r r r r r r r r r r r}
    \toprule

    & \multicolumn{11}{c}{\textcolor{dpd2}{\textbf{2007Q2 -- 2019Q4}}}
    && \multicolumn{11}{c}{\textcolor{dpd2}{\textbf{2021Q1 -- 2025Q1}}}
    \\
    \cmidrule(lr){2-12}
    \cmidrule(lr){14-24}
    & FAAR & RF & LGB & LGB+ & LGB$^{\texttt{A}}+$ & KRR & NN & HNN & RR & SPF & TPFN 
    && FAAR & RF & LGB & LGB+ & LGB$^{\texttt{A}}+$ & KRR & NN & HNN & RR & SPF & TPFN \\

    \midrule
    \addlinespace[0.5em]
    \multicolumn{24}{l}{\textcolor{dpd2}{\textbf{Panel A: Squared Error}}} \\
    \addlinespace[0.3em]

      Return
      & -0.43 & 0.07 & -0.13 & 0.19 & 0.15 & -0.05 & 0.03 & {\textbf{0.22}} & -0.11 & {\color{ForestGreen}\textbf{0.25}} & -0.05 &  & -1.04 & -0.39 & -1.48 & -0.07 & -0.29 & {\color{ForestGreen}\textbf{0.12}} & -0.66 & {\textbf{0.07}} & -1.07 & -1.54 & -0.24 \\
      \rowcolor{RowAlt}
      Sharpe
      & -0.37 & 0.12 & -0.17 & {\textbf{0.32}} & 0.17 & -0.08 & 0.03 & {\color{ForestGreen}\textbf{0.36}} & -0.13 & 0.31 & -0.04 &  & -0.47 & -0.57 & -1.02 & -0.17 & -0.54 & {\color{ForestGreen}\textbf{0.30}} & -0.48 & {\textbf{0.09}} & -0.58 & -0.87 & -0.47 \\
      Sortino
      & -0.41 & 0.25 & -0.28 & {\textbf{1.61}} & 0.39 & -0.11 & 0.05 & {\color{ForestGreen}\textbf{2.00}} & -0.18 & 1.21 & -0.10 &  & -0.48 & -0.62 & -0.94 & -0.27 & -0.61 & {\color{ForestGreen}\textbf{0.58}} & -0.50 & {\textbf{0.12}} & -0.58 & -0.84 & -0.59 \\
      \rowcolor{RowAlt}
      Omega
      & 0.33 & 1.33 & 0.67 & {\textbf{2.53}} & 1.69 & 0.79 & 1.09 & {\color{ForestGreen}\textbf{3.01}} & 0.69 & 2.38 & 0.87 &  & 0.25 & 0.37 & 0.07 & 0.76 & 0.40 & {\color{ForestGreen}\textbf{1.59}} & 0.32 & {\textbf{1.14}} & 0.26 & 0.21 & 0.50 \\
      MaxDD
      & -25.55 & -4.61 & -13.08 & {\color{ForestGreen}\textbf{-0.34}} & {\textbf{-0.94}} & -8.22 & -8.25 & -1.99 & -12.74 & -3.00 & -13.14 &  & -3.76 & -10.03 & -25.25 & -3.80 & -7.65 & {\color{ForestGreen}\textbf{-0.47}} & -15.52 & {\textbf{-0.74}} & -24.37 & -32.80 & -7.22 \\
      \rowcolor{RowAlt}
      Edge
      & 0.01 & 0.00 & 0.06 & 0.01 & 0.07 & 0.15 & 0.06 & 0.07 & 0.00 & {\textbf{0.52}} & {\color{ForestGreen}\textbf{0.86}} &  & 0.03 & 0.00 & {\textbf{0.06}} & 0.00 & 0.00 & 0.01 & 0.06 & {\color{ForestGreen}\textbf{1.83}} & 0.00 & 0.00 & 0.04 \\

    \addlinespace[0.5em]
    \midrule
    \addlinespace[0.5em]
    \multicolumn{24}{l}{\textcolor{dpd2}{\textbf{Panel B: Absolute Error}}} \\
    \addlinespace[0.3em]

      Return
      & -0.31 & -0.04 & -0.20 & 0.04 & -0.01 & -0.01 & -0.05 & {\textbf{0.04}} & -0.11 & {\color{ForestGreen}\textbf{0.09}} & -0.16 &  & -0.13 & -0.08 & -0.32 & 0.01 & -0.13 & {\textbf{0.05}} & -0.04 & {\color{ForestGreen}\textbf{0.14}} & -0.18 & -0.32 & -0.06 \\
      \rowcolor{RowAlt}
      Sharpe
      & -0.76 & -0.13 & -0.54 & 0.12 & -0.03 & -0.02 & -0.13 & {\textbf{0.17}} & -0.32 & {\color{ForestGreen}\textbf{0.21}} & -0.40 &  & -0.25 & -0.29 & -0.76 & 0.05 & -0.55 & {\textbf{0.25}} & -0.10 & {\color{ForestGreen}\textbf{0.32}} & -0.34 & -0.51 & -0.22 \\
      Sortino
      & -0.76 & -0.17 & -0.60 & 0.23 & -0.03 & -0.02 & -0.18 & {\textbf{0.27}} & -0.38 & {\color{ForestGreen}\textbf{0.37}} & -0.48 &  & -0.29 & -0.35 & -0.77 & 0.10 & -0.65 & {\textbf{0.40}} & -0.13 & {\color{ForestGreen}\textbf{0.56}} & -0.39 & -0.56 & -0.31 \\
      \rowcolor{RowAlt}
      Omega
      & 0.28 & 0.83 & 0.46 & 1.21 & 0.95 & 0.97 & 0.81 & {\textbf{1.25}} & 0.61 & {\color{ForestGreen}\textbf{1.37}} & 0.53 &  & 0.67 & 0.69 & 0.33 & 1.08 & 0.47 & {\textbf{1.35}} & 0.86 & {\color{ForestGreen}\textbf{1.48}} & 0.58 & 0.49 & 0.75 \\
      MaxDD
      & -14.62 & -4.60 & -10.57 & {\color{ForestGreen}\textbf{-0.97}} & -3.56 & -3.88 & -5.52 & {\textbf{-2.33}} & -7.21 & -3.92 & -8.83 &  & -1.30 & -3.91 & -6.55 & -0.66 & -3.37 & {\textbf{-0.45}} & -4.27 & {\color{ForestGreen}\textbf{-0.43}} & -6.95 & -10.14 & -3.40 \\
      \rowcolor{RowAlt}
      Edge
      & 0.08 & 0.01 & 0.15 & 0.05 & 0.18 & {\textbf{0.58}} & 0.10 & 0.09 & 0.01 & {\color{ForestGreen}\textbf{0.90}} & 0.24 &  & 0.15 & 0.00 & 0.20 & 0.00 & 0.00 & 0.08 & {\textbf{0.29}} & {\color{ForestGreen}\textbf{2.30}} & 0.00 & 0.00 & 0.17 \\

    \addlinespace[0.5em]
    \midrule
    \addlinespace[0.5em]
    \multicolumn{24}{l}{\textcolor{dpd2}{\textbf{Panel C: Classical Forecast Accuracy}}} \\
    \addlinespace[0.3em]

      RMSE
      & 1.19 & 0.96 & 1.06 & 0.90 & 0.92 & 1.03 & 0.99 & {\textbf{0.89}} & 1.05 & {\color{ForestGreen}\textbf{0.86}} & 1.02 &  & 1.43 & 1.18 & 1.58 & 1.03 & 1.14 & {\color{ForestGreen}\textbf{0.94}} & 1.29 & {\textbf{0.96}} & 1.44 & 1.59 & 1.11 \\
      \rowcolor{RowAlt}
      MAE
      & 1.31 & 1.04 & 1.20 & 0.96 & 1.01 & 1.01 & 1.05 & {\textbf{0.96}} & 1.11 & {\color{ForestGreen}\textbf{0.91}} & 1.16 &  & 1.13 & 1.08 & 1.32 & 0.99 & 1.13 & {\textbf{0.95}} & 1.04 & {\color{ForestGreen}\textbf{0.86}} & 1.18 & 1.32 & 1.06 \\
      $\rho(1)$
      & 0.15 & 0.20 & 0.26 & 0.10 & {\textbf{0.04}} & 0.29 & 0.08 & {\color{ForestGreen}\textbf{0.02}} & 0.16 & 0.18 & 0.21 &  & 0.32 & 0.55 & 0.46 & 0.45 & 0.49 & {\color{ForestGreen}\textbf{0.12}} & {\textbf{0.14}} & 0.16 & 0.49 & 0.82 & 0.37 \\
      \rowcolor{RowAlt}
      DM $t$-stat
      & -1.18 & 0.72 & -0.59 & 1.33 & 0.92 & -0.27 & 0.10 & {\color{ForestGreen}\textbf{1.40}} & -0.43 & {\textbf{1.38}} & -0.16 &  & -1.00 & -1.02 & -2.13 & -0.43 & -1.15 & {\color{ForestGreen}\textbf{0.62}} & -1.15 & {\textbf{0.18}} & -1.23 & -1.19 & -0.75 \\

    \addlinespace[0.3em]
    \bottomrule
  \end{tabular}%
  }

  \vspace{0.25em}
  \parbox{\linewidth}{\scriptsize
    \textit{Notes}: Panels A--B report risk-adjusted metrics; Panel C reports classical forecast accuracy metrics. $\rho(1)$ = first-order autocorrelation of errors.
    Best: \textcolor{ForestGreen}{\textbf{bold green}}; second-best: \textbf{bold}.
  }
  }

\end{table}
\end{landscape}


\begin{landscape}
\begin{table}[t!]
  \centering
  \caption{\normalsize Inflation ($h=2$)}
  \vspace*{-0.65em}
  \label{tab:infl_h2}

  {\fontfamily{phv}\selectfont
  \resizebox{\linewidth}{!}{%
  \scriptsize
  \setlength{\tabcolsep}{0.3em}
  \renewcommand{\arraystretch}{1.65}
  \begin{tabular}{l r r r r r r r r r r r c r r r r r r r r r r r}
    \toprule

    & \multicolumn{11}{c}{\textcolor{dpd2}{\textbf{2007Q2 -- 2019Q4}}}
    && \multicolumn{11}{c}{\textcolor{dpd2}{\textbf{2021Q1 -- 2025Q1}}}
    \\
    \cmidrule(lr){2-12}
    \cmidrule(lr){14-24}
    & FAAR & RF & LGB & LGB+ & LGB$^{\texttt{A}}+$ & KRR & NN & HNN & RR & SPF & TPFN 
    && FAAR & RF & LGB & LGB+ & LGB$^{\texttt{A}}+$ & KRR & NN & HNN & RR & SPF & TPFN \\

    \midrule
    \addlinespace[0.5em]
    \multicolumn{24}{l}{\textcolor{dpd2}{\textbf{Panel A: Squared Error}}} \\
    \addlinespace[0.3em]

      Return
      & -0.30 & 0.31 & -0.03 & 0.17 & 0.09 & 0.11 & {\textbf{0.36}} & 0.26 & 0.17 & {\color{ForestGreen}\textbf{0.42}} & 0.10 &  & -2.91 & -0.35 & -0.34 & -0.05 & -0.10 & {\color{ForestGreen}\textbf{0.51}} & -0.03 & 0.12 & -0.53 & -0.67 & {\textbf{0.17}} \\
      \rowcolor{RowAlt}
      Sharpe
      & -0.20 & {\textbf{0.58}} & -0.05 & 0.31 & 0.12 & 0.20 & 0.49 & 0.52 & 0.34 & {\color{ForestGreen}\textbf{0.60}} & 0.15 &  & -0.49 & -0.71 & -0.84 & -0.14 & -0.45 & {\color{ForestGreen}\textbf{1.40}} & -0.05 & 0.32 & -0.64 & -0.65 & {\textbf{0.43}} \\
      Sortino
      & -0.22 & {\color{ForestGreen}\textbf{4.62}} & -0.06 & 1.10 & 0.16 & 0.36 & 2.65 & 2.01 & 1.44 & {\textbf{4.15}} & 0.34 &  & -0.49 & -0.71 & -0.84 & -0.16 & -0.50 & {\color{ForestGreen}\textbf{9.97}} & -0.05 & 0.45 & -0.67 & -0.67 & {\textbf{0.63}} \\
      \rowcolor{RowAlt}
      Omega
      & 0.53 & {\textbf{6.29}} & 0.89 & 2.17 & 1.38 & 1.80 & 5.84 & 3.87 & 2.40 & {\color{ForestGreen}\textbf{7.12}} & 1.43 &  & 0.10 & 0.30 & 0.32 & 0.81 & 0.53 & {\color{ForestGreen}\textbf{11.95}} & 0.91 & 1.50 & 0.34 & 0.32 & {\textbf{1.88}} \\
      MaxDD
      & -25.56 & -0.67 & -10.61 & {\color{ForestGreen}\textbf{-0.49}} & -8.10 & -5.04 & -1.84 & -2.74 & {\textbf{-0.57}} & -2.46 & -6.23 &  & -3.36 & -7.67 & -4.80 & -4.02 & -3.09 & {\color{ForestGreen}\textbf{-0.08}} & -4.36 & {\textbf{-0.58}} & -12.64 & -16.40 & -3.13 \\
      \rowcolor{RowAlt}
      Edge
      & 0.07 & 0.07 & 0.01 & 0.07 & 0.00 & {\textbf{0.13}} & {\color{ForestGreen}\textbf{0.87}} & 0.11 & 0.05 & 0.12 & 0.03 &  & 0.13 & 0.00 & 0.00 & {\textbf{0.34}} & 0.00 & {\color{ForestGreen}\textbf{3.10}} & 0.12 & 0.09 & 0.00 & 0.12 & 0.00 \\

    \addlinespace[0.5em]
    \midrule
    \addlinespace[0.5em]
    \multicolumn{24}{l}{\textcolor{dpd2}{\textbf{Panel B: Absolute Error}}} \\
    \addlinespace[0.3em]

      Return
      & -0.13 & 0.17 & 0.01 & 0.08 & 0.03 & 0.13 & {\textbf{0.21}} & 0.13 & 0.05 & {\color{ForestGreen}\textbf{0.26}} & 0.03 &  & -0.42 & -0.07 & -0.11 & 0.04 & 0.03 & {\color{ForestGreen}\textbf{0.28}} & 0.13 & 0.13 & -0.05 & -0.04 & {\textbf{0.20}} \\
      \rowcolor{RowAlt}
      Sharpe
      & -0.25 & 0.68 & 0.03 & 0.26 & 0.11 & 0.44 & {\color{ForestGreen}\textbf{0.82}} & 0.51 & 0.19 & {\textbf{0.80}} & 0.10 &  & -0.47 & -0.30 & -0.45 & 0.26 & 0.19 & {\color{ForestGreen}\textbf{1.20}} & 0.53 & 0.49 & -0.15 & -0.09 & {\textbf{0.89}} \\
      Sortino
      & -0.28 & 1.97 & 0.04 & 0.59 & 0.15 & 1.24 & {\color{ForestGreen}\textbf{2.27}} & 0.90 & 0.37 & {\textbf{2.14}} & 0.15 &  & -0.49 & -0.36 & -0.54 & 0.38 & 0.29 & {\color{ForestGreen}\textbf{3.81}} & 0.89 & 0.77 & -0.20 & -0.13 & {\textbf{1.85}} \\
      \rowcolor{RowAlt}
      Omega
      & 0.61 & 2.83 & 1.05 & 1.48 & 1.22 & 2.25 & {\color{ForestGreen}\textbf{3.84}} & 2.02 & 1.30 & {\textbf{3.72}} & 1.15 &  & 0.30 & 0.69 & 0.58 & 1.39 & 1.24 & {\color{ForestGreen}\textbf{5.09}} & 2.02 & 1.78 & 0.82 & 0.88 & {\textbf{3.02}} \\
      MaxDD
      & -8.26 & -0.91 & -4.77 & {\textbf{-0.65}} & -2.89 & -1.71 & -1.82 & -3.15 & {\color{ForestGreen}\textbf{-0.64}} & -3.02 & -3.08 &  & -1.93 & -2.58 & -1.83 & -0.66 & -0.91 & {\color{ForestGreen}\textbf{-0.17}} & -0.65 & {\textbf{-0.37}} & -3.99 & -5.21 & -0.66 \\
      \rowcolor{RowAlt}
      Edge
      & 0.20 & 0.08 & 0.10 & 0.23 & 0.03 & 0.32 & {\color{ForestGreen}\textbf{0.39}} & 0.17 & 0.20 & {\textbf{0.35}} & 0.17 &  & 0.33 & 0.01 & 0.00 & 0.33 & 0.03 & {\color{ForestGreen}\textbf{2.26}} & 0.24 & 0.32 & 0.00 & {\textbf{0.64}} & 0.08 \\

    \addlinespace[0.5em]
    \midrule
    \addlinespace[0.5em]
    \multicolumn{24}{l}{\textcolor{dpd2}{\textbf{Panel C: Classical Forecast Accuracy}}} \\
    \addlinespace[0.3em]

      RMSE
      & 1.14 & 0.83 & 1.01 & 0.91 & 0.95 & 0.94 & {\textbf{0.80}} & 0.86 & 0.91 & {\color{ForestGreen}\textbf{0.76}} & 0.95 &  & 1.98 & 1.16 & 1.16 & 1.02 & 1.05 & {\color{ForestGreen}\textbf{0.70}} & 1.01 & 0.94 & 1.24 & 1.29 & {\textbf{0.91}} \\
      \rowcolor{RowAlt}
      MAE
      & 1.13 & 0.83 & 0.99 & 0.92 & 0.97 & 0.87 & {\textbf{0.79}} & 0.87 & 0.95 & {\color{ForestGreen}\textbf{0.74}} & 0.97 &  & 1.42 & 1.07 & 1.11 & 0.96 & 0.97 & {\color{ForestGreen}\textbf{0.72}} & 0.87 & 0.87 & 1.05 & 1.04 & {\textbf{0.80}} \\
      $\rho(1)$
      & 0.27 & 0.31 & 0.37 & 0.34 & 0.27 & 0.37 & {\textbf{0.25}} & 0.27 & 0.39 & {\color{ForestGreen}\textbf{0.24}} & 0.34 &  & {\color{ForestGreen}\textbf{0.22}} & 0.81 & 0.72 & 0.75 & 0.75 & {\textbf{0.45}} & 0.58 & 0.50 & 0.81 & 0.87 & 0.70 \\
      \rowcolor{RowAlt}
      DM $t$-stat
      & -0.67 & {\textbf{1.65}} & -0.16 & 1.11 & 0.51 & 0.65 & 1.62 & 1.54 & 1.02 & {\color{ForestGreen}\textbf{1.71}} & 0.59 &  & -1.01 & -1.04 & -1.72 & -0.24 & -0.94 & {\color{ForestGreen}\textbf{1.48}} & -0.11 & 0.58 & -0.85 & -0.99 & {\textbf{0.73}} \\

    \addlinespace[0.3em]
    \bottomrule
  \end{tabular}%
  }

  \vspace{0.25em}
  \parbox{\linewidth}{\scriptsize
    \textit{Notes}: Panels A--B report risk-adjusted metrics; Panel C reports classical forecast accuracy metrics. $\rho(1)$ = first-order autocorrelation of errors.
    Best: \textcolor{ForestGreen}{\textbf{bold green}}; second-best: \textbf{bold}.
  }
  }

\end{table}
\end{landscape}


\begin{landscape}
\begin{table}[t!]
  \centering
  \caption{\normalsize Inflation ($h=4$)}
  \vspace*{-0.65em}
  \label{tab:infl_h4}

  {\fontfamily{phv}\selectfont
  \resizebox{\linewidth}{!}{%
  \scriptsize
  \setlength{\tabcolsep}{0.3em}
  \renewcommand{\arraystretch}{1.65}
  \begin{tabular}{l r r r r r r r r r r r c r r r r r r r r r r r}
    \toprule

    & \multicolumn{11}{c}{\textcolor{dpd2}{\textbf{2007Q2 -- 2019Q4}}}
    && \multicolumn{11}{c}{\textcolor{dpd2}{\textbf{2021Q1 -- 2025Q1}}}
    \\
    \cmidrule(lr){2-12}
    \cmidrule(lr){14-24}
    & FAAR & RF & LGB & LGB+ & LGB$^{\texttt{A}}+$ & KRR & NN & HNN & RR & SPF & TPFN 
    && FAAR & RF & LGB & LGB+ & LGB$^{\texttt{A}}+$ & KRR & NN & HNN & RR & SPF & TPFN \\

    \midrule
    \addlinespace[0.5em]
    \multicolumn{24}{l}{\textcolor{dpd2}{\textbf{Panel A: Squared Error}}} \\
    \addlinespace[0.3em]

      Return
      & -1.15 & 0.18 & 0.24 & 0.18 & 0.24 & 0.22 & 0.24 & {\textbf{0.29}} & 0.07 & {\color{ForestGreen}\textbf{0.36}} & -0.26 &  & -0.10 & -0.50 & -0.56 & -0.40 & -0.43 & {\color{ForestGreen}\textbf{0.42}} & -0.14 & -0.53 & -0.59 & 0.00 & {\textbf{0.08}} \\
      \rowcolor{RowAlt}
      Sharpe
      & -0.39 & 0.41 & 0.46 & 0.41 & 0.43 & 0.39 & {\textbf{0.48}} & 0.46 & 0.15 & {\color{ForestGreen}\textbf{0.54}} & -0.29 &  & -0.19 & -1.37 & -0.69 & -0.84 & -0.86 & {\color{ForestGreen}\textbf{0.94}} & -0.21 & -0.41 & -0.58 & 0.01 & {\textbf{0.23}} \\
      Sortino
      & -0.39 & 1.28 & {\textbf{1.91}} & 1.45 & 1.17 & 0.97 & 1.47 & 1.42 & 0.30 & {\color{ForestGreen}\textbf{2.85}} & -0.34 &  & -0.30 & -1.16 & -0.79 & -0.92 & -0.82 & {\color{ForestGreen}\textbf{8.69}} & -0.24 & -0.43 & -0.59 & 0.01 & {\textbf{0.40}} \\
      \rowcolor{RowAlt}
      Omega
      & 0.12 & 2.38 & 3.20 & 2.62 & 2.63 & 2.53 & {\textbf{3.21}} & 2.74 & 1.39 & {\color{ForestGreen}\textbf{4.74}} & 0.52 &  & 0.74 & 0.06 & 0.38 & 0.31 & 0.16 & {\color{ForestGreen}\textbf{8.36}} & 0.67 & 0.38 & 0.30 & 1.01 & {\textbf{1.40}} \\
      MaxDD
      & -60.05 & {\textbf{-0.42}} & -1.00 & {\color{ForestGreen}\textbf{-0.25}} & -4.41 & -0.78 & -0.56 & -3.57 & -0.58 & -2.19 & -21.50 &  & -4.55 & -7.55 & -13.52 & -8.16 & -7.71 & {\color{ForestGreen}\textbf{-0.09}} & -6.81 & -11.22 & -14.19 & -5.13 & {\textbf{-3.12}} \\
      \rowcolor{RowAlt}
      Edge
      & 0.07 & 0.00 & 0.03 & 0.11 & 0.07 & 0.07 & 0.00 & {\textbf{0.57}} & 0.00 & {\color{ForestGreen}\textbf{1.37}} & 0.10 &  & {\textbf{0.50}} & 0.00 & 0.04 & 0.00 & 0.00 & {\color{ForestGreen}\textbf{2.78}} & 0.03 & 0.09 & 0.01 & 0.10 & 0.01 \\

    \addlinespace[0.5em]
    \midrule
    \addlinespace[0.5em]
    \multicolumn{24}{l}{\textcolor{dpd2}{\textbf{Panel B: Absolute Error}}} \\
    \addlinespace[0.3em]

      Return
      & -0.33 & 0.06 & 0.09 & 0.06 & 0.12 & 0.11 & 0.12 & {\textbf{0.12}} & -0.06 & {\color{ForestGreen}\textbf{0.18}} & -0.22 &  & -0.14 & -0.31 & -0.36 & -0.29 & -0.18 & {\color{ForestGreen}\textbf{0.14}} & 0.08 & -0.06 & -0.04 & 0.09 & {\textbf{0.11}} \\
      \rowcolor{RowAlt}
      Sharpe
      & -0.44 & 0.17 & 0.26 & 0.18 & {\textbf{0.38}} & 0.30 & 0.37 & 0.29 & -0.21 & {\color{ForestGreen}\textbf{0.51}} & -0.49 &  & -0.52 & -1.68 & -0.87 & -1.18 & -0.85 & {\color{ForestGreen}\textbf{0.59}} & 0.23 & -0.14 & -0.09 & 0.25 & {\textbf{0.37}} \\
      Sortino
      & -0.45 & 0.28 & 0.48 & 0.35 & 0.64 & 0.45 & {\textbf{0.70}} & 0.52 & -0.27 & {\color{ForestGreen}\textbf{1.09}} & -0.52 &  & -0.73 & -1.31 & -1.00 & -1.19 & -0.87 & {\color{ForestGreen}\textbf{1.66}} & 0.38 & -0.18 & -0.12 & 0.41 & {\textbf{0.77}} \\
      \rowcolor{RowAlt}
      Omega
      & 0.34 & 1.28 & 1.47 & 1.33 & 1.68 & 1.58 & {\textbf{1.75}} & 1.46 & 0.72 & {\color{ForestGreen}\textbf{2.11}} & 0.49 &  & 0.49 & 0.06 & 0.35 & 0.24 & 0.28 & {\color{ForestGreen}\textbf{2.43}} & 1.40 & 0.83 & 0.88 & 1.38 & {\textbf{1.66}} \\
      MaxDD
      & -17.35 & {\textbf{-0.55}} & -4.05 & {\color{ForestGreen}\textbf{-0.54}} & -4.01 & -0.79 & -0.58 & -3.35 & -7.50 & -2.48 & -13.58 &  & -3.54 & -4.52 & -8.34 & -5.55 & -2.98 & {\color{ForestGreen}\textbf{-0.29}} & -2.38 & -2.21 & -4.51 & -2.78 & {\textbf{-2.00}} \\
      \rowcolor{RowAlt}
      Edge
      & 0.22 & 0.00 & 0.17 & 0.18 & {\textbf{0.25}} & 0.17 & 0.00 & {\color{ForestGreen}\textbf{0.75}} & 0.01 & 0.21 & 0.18 &  & 0.37 & 0.00 & 0.18 & 0.00 & 0.00 & {\color{ForestGreen}\textbf{0.99}} & 0.19 & 0.29 & 0.06 & {\textbf{0.59}} & 0.05 \\

    \addlinespace[0.5em]
    \midrule
    \addlinespace[0.5em]
    \multicolumn{24}{l}{\textcolor{dpd2}{\textbf{Panel C: Classical Forecast Accuracy}}} \\
    \addlinespace[0.3em]

      RMSE
      & 1.46 & 0.90 & 0.87 & 0.91 & 0.87 & 0.88 & 0.87 & {\textbf{0.84}} & 0.96 & {\color{ForestGreen}\textbf{0.80}} & 1.12 &  & 1.05 & 1.22 & 1.25 & 1.18 & 1.20 & {\color{ForestGreen}\textbf{0.76}} & 1.07 & 1.24 & 1.26 & 1.00 & {\textbf{0.96}} \\
      \rowcolor{RowAlt}
      MAE
      & 1.33 & 0.94 & 0.91 & 0.94 & 0.88 & 0.89 & 0.88 & {\textbf{0.88}} & 1.06 & {\color{ForestGreen}\textbf{0.82}} & 1.22 &  & 1.14 & 1.31 & 1.36 & 1.29 & 1.18 & {\color{ForestGreen}\textbf{0.86}} & 0.92 & 1.06 & 1.04 & 0.91 & {\textbf{0.89}} \\
      $\rho(1)$
      & 0.39 & 0.37 & 0.35 & 0.38 & 0.30 & 0.39 & 0.29 & {\color{ForestGreen}\textbf{0.22}} & 0.41 & {\textbf{0.24}} & 0.44 &  & {\textbf{0.57}} & 0.87 & 0.70 & 0.81 & 0.79 & 0.79 & 0.69 & {\color{ForestGreen}\textbf{0.42}} & 0.76 & 0.87 & 0.86 \\
      \rowcolor{RowAlt}
      DM $t$-stat
      & -1.12 & 1.52 & 1.63 & 1.49 & 1.44 & 1.49 & {\textbf{1.71}} & 1.66 & 0.49 & {\color{ForestGreen}\textbf{1.91}} & -0.96 &  & -0.47 & -2.84 & -1.43 & -1.49 & -1.71 & {\color{ForestGreen}\textbf{1.79}} & -0.45 & -0.88 & -0.95 & 0.01 & {\textbf{0.35}} \\

    \addlinespace[0.3em]
    \bottomrule
  \end{tabular}%
  }

  \vspace{0.25em}
  \parbox{\linewidth}{\scriptsize
    \textit{Notes}: Panels A--B report risk-adjusted metrics; Panel C reports classical forecast accuracy metrics. $\rho(1)$ = first-order autocorrelation of errors.
    Best: \textcolor{ForestGreen}{\textbf{bold green}}; second-best: \textbf{bold}.
  }
  }

\end{table}
\end{landscape}


\begin{landscape}
\begin{table}[t!]
  \centering
  \caption{\normalsize Housing Starts ($h=1$)}
  \vspace*{-0.65em}
  \label{tab:houst_h1}

  {\fontfamily{phv}\selectfont
  \resizebox{\linewidth}{!}{%
  \scriptsize
  \setlength{\tabcolsep}{0.3em}
  \renewcommand{\arraystretch}{1.65}
  \begin{tabular}{l r r r r r r r r r r c r r r r r r r r r r}
    \toprule

    & \multicolumn{10}{c}{\textcolor{dpd2}{\textbf{2007Q2 -- 2019Q4}}}
    && \multicolumn{10}{c}{\textcolor{dpd2}{\textbf{2021Q1 -- 2025Q1}}}
    \\
    \cmidrule(lr){2-11}
    \cmidrule(lr){13-22}
    & FAAR & RF & LGB & LGB+ & LGB$^{\texttt{A}}+$ & KRR & NN & RR & SPF & TPFN 
    && FAAR & RF & LGB & LGB+ & LGB$^{\texttt{A}}+$ & KRR & NN & RR & SPF & TPFN \\

    \midrule
    \addlinespace[0.5em]
    \multicolumn{22}{l}{\textcolor{dpd2}{\textbf{Panel A: Squared Error}}} \\
    \addlinespace[0.3em]

      Return
      & -1.05 & -0.03 & -0.17 & {\color{ForestGreen}\textbf{0.06}} & {\textbf{0.00}} & -0.12 & -0.27 & -0.17 & -0.04 & -0.06 &  & -4.87 & 0.17 & 0.09 & 0.08 & {\color{ForestGreen}\textbf{0.31}} & 0.12 & -0.29 & 0.20 & {\textbf{0.30}} & -0.19 \\
      \rowcolor{RowAlt}
      Sharpe
      & -0.69 & -0.04 & -0.19 & {\color{ForestGreen}\textbf{0.09}} & {\textbf{0.00}} & -0.17 & -0.27 & -0.21 & -0.08 & -0.08 &  & -1.04 & 0.32 & 0.13 & 0.11 & {\textbf{0.48}} & 0.18 & -0.33 & 0.43 & {\color{ForestGreen}\textbf{0.83}} & -0.18 \\
      Sortino
      & -0.67 & -0.04 & -0.20 & {\color{ForestGreen}\textbf{0.10}} & {\textbf{0.00}} & -0.18 & -0.28 & -0.22 & -0.11 & -0.08 &  & -0.94 & 0.55 & 0.17 & 0.18 & {\textbf{0.95}} & 0.30 & -0.43 & 0.81 & {\color{ForestGreen}\textbf{2.13}} & -0.22 \\
      \rowcolor{RowAlt}
      Omega
      & 0.12 & 0.89 & 0.63 & {\color{ForestGreen}\textbf{1.23}} & {\textbf{1.01}} & 0.67 & 0.49 & 0.55 & 0.87 & 0.81 &  & 0.03 & 1.61 & 1.21 & 1.19 & {\textbf{2.02}} & 1.29 & 0.64 & 1.89 & {\color{ForestGreen}\textbf{4.02}} & 0.72 \\
      MaxDD
      & -53.41 & -12.61 & -15.76 & -7.83 & {\color{ForestGreen}\textbf{-7.19}} & -13.69 & -21.88 & -14.85 & {\textbf{-7.34}} & -13.39 &  & -67.71 & -0.67 & -0.78 & -0.76 & {\textbf{-0.53}} & -4.92 & -7.63 & -3.22 & {\color{ForestGreen}\textbf{-0.31}} & -8.35 \\
      \rowcolor{RowAlt}
      Edge
      & 0.03 & 0.00 & {\textbf{0.23}} & 0.03 & 0.06 & 0.01 & 0.07 & 0.04 & {\color{ForestGreen}\textbf{1.00}} & 0.07 &  & 0.01 & 0.00 & {\textbf{0.20}} & 0.06 & 0.01 & 0.01 & 0.19 & 0.03 & {\color{ForestGreen}\textbf{0.49}} & 0.00 \\

    \addlinespace[0.5em]
    \midrule
    \addlinespace[0.5em]
    \multicolumn{22}{l}{\textcolor{dpd2}{\textbf{Panel B: Absolute Error}}} \\
    \addlinespace[0.3em]

      Return
      & -0.40 & {\color{ForestGreen}\textbf{0.14}} & 0.06 & {\textbf{0.13}} & 0.07 & 0.01 & -0.03 & 0.01 & 0.05 & 0.08 &  & -1.15 & 0.13 & 0.12 & 0.08 & {\color{ForestGreen}\textbf{0.18}} & 0.05 & -0.09 & 0.08 & {\textbf{0.18}} & -0.08 \\
      \rowcolor{RowAlt}
      Sharpe
      & -0.98 & {\textbf{0.51}} & 0.21 & {\color{ForestGreen}\textbf{0.52}} & 0.33 & 0.03 & -0.09 & 0.05 & 0.17 & 0.30 &  & -1.32 & 0.43 & 0.32 & 0.19 & {\textbf{0.44}} & 0.14 & -0.18 & 0.30 & {\color{ForestGreen}\textbf{0.79}} & -0.16 \\
      Sortino
      & -0.93 & {\textbf{0.75}} & 0.28 & {\color{ForestGreen}\textbf{0.88}} & 0.52 & 0.04 & -0.11 & 0.06 & 0.26 & 0.41 &  & -1.13 & {\textbf{0.82}} & 0.50 & 0.33 & 0.78 & 0.22 & -0.25 & 0.50 & {\color{ForestGreen}\textbf{2.38}} & -0.20 \\
      \rowcolor{RowAlt}
      Omega
      & 0.22 & {\color{ForestGreen}\textbf{2.16}} & 1.35 & {\textbf{2.13}} & 1.61 & 1.05 & 0.89 & 1.07 & 1.24 & 1.56 &  & 0.09 & {\textbf{1.78}} & 1.48 & 1.32 & 1.78 & 1.18 & 0.80 & 1.50 & {\color{ForestGreen}\textbf{3.39}} & 0.80 \\
      MaxDD
      & -19.40 & -2.77 & -2.84 & {\color{ForestGreen}\textbf{-0.85}} & {\textbf{-2.19}} & -4.46 & -7.31 & -4.13 & -3.56 & -3.98 &  & -16.49 & -0.55 & {\textbf{-0.50}} & -0.61 & -0.51 & -3.21 & -3.67 & -2.38 & {\color{ForestGreen}\textbf{-0.30}} & -3.98 \\
      \rowcolor{RowAlt}
      Edge
      & 0.05 & 0.07 & {\textbf{0.43}} & 0.13 & 0.07 & 0.07 & 0.16 & 0.07 & {\color{ForestGreen}\textbf{1.59}} & 0.15 &  & 0.07 & 0.00 & 0.32 & 0.15 & 0.06 & 0.09 & {\color{ForestGreen}\textbf{0.49}} & 0.10 & {\textbf{0.46}} & 0.00 \\

    \addlinespace[0.5em]
    \midrule
    \addlinespace[0.5em]
    \multicolumn{22}{l}{\textcolor{dpd2}{\textbf{Panel C: Classical Forecast Accuracy}}} \\
    \addlinespace[0.3em]

      RMSE
      & 1.43 & 1.02 & 1.08 & {\color{ForestGreen}\textbf{0.97}} & {\textbf{1.00}} & 1.06 & 1.12 & 1.08 & 1.02 & 1.03 &  & 2.42 & 0.91 & 0.95 & 0.96 & {\color{ForestGreen}\textbf{0.83}} & 0.94 & 1.13 & 0.89 & {\textbf{0.84}} & 1.09 \\
      \rowcolor{RowAlt}
      MAE
      & 1.40 & {\color{ForestGreen}\textbf{0.86}} & 0.94 & {\textbf{0.87}} & 0.93 & 0.99 & 1.03 & 0.99 & 0.95 & 0.92 &  & 2.15 & 0.87 & 0.88 & 0.92 & {\color{ForestGreen}\textbf{0.82}} & 0.95 & 1.09 & 0.92 & {\textbf{0.82}} & 1.08 \\
      $\rho(1)$
      & 0.48 & 0.30 & 0.26 & 0.24 & 0.26 & 0.37 & {\textbf{0.23}} & 0.41 & 0.34 & {\color{ForestGreen}\textbf{0.23}} &  & 0.34 & -0.14 & -0.26 & {\textbf{-0.09}} & {\color{ForestGreen}\textbf{-0.04}} & 0.11 & 0.14 & 0.09 & 0.17 & 0.21 \\
      \rowcolor{RowAlt}
      DM $t$-stat
      & -2.46 & -0.14 & -0.67 & {\color{ForestGreen}\textbf{0.36}} & {\textbf{0.01}} & -0.60 & -0.98 & -0.69 & -0.31 & -0.28 &  & -1.88 & 0.61 & 0.26 & 0.21 & {\textbf{1.02}} & 0.33 & -0.66 & 0.79 & {\color{ForestGreen}\textbf{1.87}} & -0.37 \\

    \addlinespace[0.3em]
    \bottomrule
  \end{tabular}%
  }

  \vspace{0.25em}
  \parbox{\linewidth}{\scriptsize
    \textit{Notes}: Panels A--B report risk-adjusted metrics; Panel C reports classical forecast accuracy metrics. $\rho(1)$ = first-order autocorrelation of errors.
    Best: \textcolor{ForestGreen}{\textbf{bold green}}; second-best: \textbf{bold}.
  }
  }

\end{table}
\end{landscape}


\begin{landscape}
\begin{table}[t!]
  \centering
  \caption{\normalsize Housing Starts ($h=2$)}
  \vspace*{-0.65em}
  \label{tab:houst_h2}

  {\fontfamily{phv}\selectfont
  \resizebox{\linewidth}{!}{%
  \scriptsize
  \setlength{\tabcolsep}{0.3em}
  \renewcommand{\arraystretch}{1.65}
  \begin{tabular}{l r r r r r r r r r r c r r r r r r r r r r}
    \toprule

    & \multicolumn{10}{c}{\textcolor{dpd2}{\textbf{2007Q2 -- 2019Q4}}}
    && \multicolumn{10}{c}{\textcolor{dpd2}{\textbf{2021Q1 -- 2025Q1}}}
    \\
    \cmidrule(lr){2-11}
    \cmidrule(lr){13-22}
    & FAAR & RF & LGB & LGB+ & LGB$^{\texttt{A}}+$ & KRR & NN & RR & SPF & TPFN 
    && FAAR & RF & LGB & LGB+ & LGB$^{\texttt{A}}+$ & KRR & NN & RR & SPF & TPFN \\

    \midrule
    \addlinespace[0.5em]
    \multicolumn{22}{l}{\textcolor{dpd2}{\textbf{Panel A: Squared Error}}} \\
    \addlinespace[0.3em]

      Return
      & -0.18 & -0.06 & -0.06 & -0.09 & -0.07 & {\color{ForestGreen}\textbf{-0.04}} & -0.23 & {\textbf{-0.05}} & -0.06 & -0.05 &  & -0.02 & {\textbf{0.09}} & -0.22 & -0.28 & -0.33 & 0.04 & -0.94 & {\color{ForestGreen}\textbf{0.13}} & 0.02 & 0.07 \\
      \rowcolor{RowAlt}
      Sharpe
      & -0.25 & -0.13 & -0.13 & -0.15 & -0.16 & {\color{ForestGreen}\textbf{-0.09}} & -0.36 & -0.12 & -0.12 & {\textbf{-0.11}} &  & -0.02 & {\textbf{0.18}} & -0.46 & -0.30 & -0.47 & 0.05 & -0.48 & {\color{ForestGreen}\textbf{0.23}} & 0.05 & 0.11 \\
      Sortino
      & -0.28 & -0.14 & -0.15 & -0.17 & -0.18 & {\color{ForestGreen}\textbf{-0.12}} & -0.39 & -0.14 & -0.16 & {\textbf{-0.13}} &  & -0.03 & {\textbf{0.27}} & -0.54 & -0.36 & -0.55 & 0.07 & -0.51 & {\color{ForestGreen}\textbf{0.34}} & 0.06 & 0.16 \\
      \rowcolor{RowAlt}
      Omega
      & 0.62 & 0.78 & 0.81 & 0.78 & 0.76 & {\color{ForestGreen}\textbf{0.87}} & 0.50 & 0.80 & {\textbf{0.83}} & 0.83 &  & 0.97 & {\textbf{1.27}} & 0.48 & 0.63 & 0.53 & 1.08 & 0.39 & {\color{ForestGreen}\textbf{1.43}} & 1.07 & 1.17 \\
      MaxDD
      & -17.66 & -10.26 & -9.81 & -16.55 & -10.78 & {\textbf{-8.82}} & -18.99 & {\color{ForestGreen}\textbf{-7.92}} & -9.64 & -9.77 &  & -7.28 & {\textbf{-0.81}} & -5.89 & -9.67 & -8.42 & -6.89 & -22.55 & -3.87 & {\color{ForestGreen}\textbf{-0.79}} & -4.43 \\
      \rowcolor{RowAlt}
      Edge
      & 0.29 & 0.03 & 0.02 & {\color{ForestGreen}\textbf{0.73}} & 0.01 & 0.07 & 0.00 & 0.00 & {\textbf{0.48}} & 0.00 &  & {\color{ForestGreen}\textbf{0.29}} & 0.00 & 0.00 & {\textbf{0.17}} & 0.07 & 0.00 & 0.12 & 0.00 & 0.01 & 0.01 \\

    \addlinespace[0.5em]
    \midrule
    \addlinespace[0.5em]
    \multicolumn{22}{l}{\textcolor{dpd2}{\textbf{Panel B: Absolute Error}}} \\
    \addlinespace[0.3em]

      Return
      & 0.02 & {\textbf{0.04}} & -0.03 & 0.00 & 0.03 & -0.01 & -0.06 & 0.02 & {\color{ForestGreen}\textbf{0.07}} & 0.01 &  & 0.03 & -0.02 & -0.10 & -0.12 & -0.21 & {\color{ForestGreen}\textbf{0.03}} & -0.22 & {\textbf{0.03}} & -0.01 & 0.02 \\
      \rowcolor{RowAlt}
      Sharpe
      & 0.06 & {\textbf{0.17}} & -0.11 & 0.01 & 0.10 & -0.05 & -0.23 & 0.08 & {\color{ForestGreen}\textbf{0.21}} & 0.05 &  & 0.05 & -0.07 & -0.33 & -0.25 & -0.50 & {\textbf{0.07}} & -0.29 & {\color{ForestGreen}\textbf{0.09}} & -0.05 & 0.06 \\
      Sortino
      & 0.08 & {\textbf{0.26}} & -0.14 & 0.02 & 0.14 & -0.07 & -0.29 & 0.13 & {\color{ForestGreen}\textbf{0.34}} & 0.08 &  & 0.07 & -0.11 & -0.45 & -0.32 & -0.60 & {\textbf{0.10}} & -0.34 & {\color{ForestGreen}\textbf{0.13}} & -0.08 & 0.08 \\
      \rowcolor{RowAlt}
      Omega
      & 1.08 & {\textbf{1.29}} & 0.86 & 1.02 & 1.15 & 0.94 & 0.73 & 1.12 & {\color{ForestGreen}\textbf{1.32}} & 1.07 &  & 1.08 & 0.92 & 0.62 & 0.71 & 0.53 & {\textbf{1.09}} & 0.64 & {\color{ForestGreen}\textbf{1.13}} & 0.93 & 1.08 \\
      MaxDD
      & -5.50 & -3.71 & -4.74 & -7.86 & -6.02 & -4.83 & -7.08 & {\color{ForestGreen}\textbf{-2.91}} & {\textbf{-3.44}} & -4.79 &  & -3.96 & {\color{ForestGreen}\textbf{-0.76}} & -3.44 & -5.23 & -5.02 & -4.13 & -7.83 & -2.61 & {\textbf{-0.81}} & -2.86 \\
      \rowcolor{RowAlt}
      Edge
      & 0.41 & 0.13 & 0.11 & {\textbf{0.70}} & 0.08 & 0.15 & 0.01 & 0.03 & {\color{ForestGreen}\textbf{0.99}} & 0.01 &  & {\color{ForestGreen}\textbf{0.45}} & 0.00 & 0.00 & {\textbf{0.40}} & 0.09 & 0.06 & 0.28 & 0.00 & 0.08 & 0.02 \\

    \addlinespace[0.5em]
    \midrule
    \addlinespace[0.5em]
    \multicolumn{22}{l}{\textcolor{dpd2}{\textbf{Panel C: Classical Forecast Accuracy}}} \\
    \addlinespace[0.3em]

      RMSE
      & 1.09 & 1.03 & 1.03 & 1.04 & 1.04 & {\color{ForestGreen}\textbf{1.02}} & 1.11 & {\textbf{1.02}} & 1.03 & 1.02 &  & 1.01 & {\textbf{0.96}} & 1.10 & 1.13 & 1.15 & 0.98 & 1.39 & {\color{ForestGreen}\textbf{0.94}} & 0.99 & 0.96 \\
      \rowcolor{RowAlt}
      MAE
      & 0.98 & {\textbf{0.96}} & 1.03 & 1.00 & 0.97 & 1.01 & 1.06 & 0.98 & {\color{ForestGreen}\textbf{0.93}} & 0.99 &  & 0.97 & 1.02 & 1.10 & 1.12 & 1.21 & {\color{ForestGreen}\textbf{0.97}} & 1.22 & {\textbf{0.97}} & 1.01 & 0.98 \\
      $\rho(1)$
      & 0.50 & 0.42 & 0.44 & 0.49 & 0.51 & {\color{ForestGreen}\textbf{0.41}} & 0.45 & 0.48 & 0.43 & {\textbf{0.42}} &  & 0.28 & 0.17 & 0.15 & 0.34 & 0.34 & 0.12 & {\color{ForestGreen}\textbf{0.06}} & {\textbf{0.07}} & 0.17 & 0.11 \\
      \rowcolor{RowAlt}
      DM $t$-stat
      & -0.69 & -0.43 & -0.46 & -0.46 & -0.42 & {\color{ForestGreen}\textbf{-0.30}} & -0.96 & -0.40 & -0.44 & {\textbf{-0.37}} &  & -0.04 & {\textbf{0.40}} & -1.00 & -0.54 & -0.97 & 0.09 & -1.05 & {\color{ForestGreen}\textbf{0.42}} & 0.10 & 0.21 \\

    \addlinespace[0.3em]
    \bottomrule
  \end{tabular}%
  }

  \vspace{0.25em}
  \parbox{\linewidth}{\scriptsize
    \textit{Notes}: Panels A--B report risk-adjusted metrics; Panel C reports classical forecast accuracy metrics. $\rho(1)$ = first-order autocorrelation of errors.
    Best: \textcolor{ForestGreen}{\textbf{bold green}}; second-best: \textbf{bold}.
  }
  }

\end{table}
\end{landscape}


\begin{landscape}
\begin{table}[t!]
  \centering
  \caption{\normalsize Housing Starts ($h=4$)}
  \vspace*{-0.65em}
  \label{tab:houst_h4}

  {\fontfamily{phv}\selectfont
  \resizebox{\linewidth}{!}{%
  \scriptsize
  \setlength{\tabcolsep}{0.3em}
  \renewcommand{\arraystretch}{1.65}
  \begin{tabular}{l r r r r r r r r r r c r r r r r r r r r r}
    \toprule

    & \multicolumn{10}{c}{\textcolor{dpd2}{\textbf{2007Q2 -- 2019Q4}}}
    && \multicolumn{10}{c}{\textcolor{dpd2}{\textbf{2021Q1 -- 2025Q1}}}
    \\
    \cmidrule(lr){2-11}
    \cmidrule(lr){13-22}
    & FAAR & RF & LGB & LGB+ & LGB$^{\texttt{A}}+$ & KRR & NN & RR & SPF & TPFN 
    && FAAR & RF & LGB & LGB+ & LGB$^{\texttt{A}}+$ & KRR & NN & RR & SPF & TPFN \\

    \midrule
    \addlinespace[0.5em]
    \multicolumn{22}{l}{\textcolor{dpd2}{\textbf{Panel A: Squared Error}}} \\
    \addlinespace[0.3em]

      Return
      & -0.11 & 0.01 & -0.07 & -0.10 & -0.04 & {\color{ForestGreen}\textbf{0.04}} & -0.09 & -0.00 & -0.08 & {\textbf{0.04}} &  & -2.23 & -0.28 & -1.09 & -0.62 & -1.11 & {\color{ForestGreen}\textbf{0.01}} & -0.94 & {\textbf{-0.02}} & -0.31 & -0.23 \\
      \rowcolor{RowAlt}
      Sharpe
      & -0.30 & 0.03 & -0.17 & -0.23 & -0.10 & {\color{ForestGreen}\textbf{0.39}} & -0.24 & -0.02 & -0.17 & {\textbf{0.14}} &  & -0.66 & -0.43 & -0.91 & -0.80 & -0.89 & {\color{ForestGreen}\textbf{0.02}} & -1.09 & {\textbf{-0.05}} & -0.74 & -0.64 \\
      Sortino
      & -0.33 & 0.04 & -0.20 & -0.25 & -0.12 & {\color{ForestGreen}\textbf{0.83}} & -0.25 & -0.02 & -0.21 & {\textbf{0.18}} &  & -0.64 & -0.50 & -0.88 & -0.81 & -0.86 & {\color{ForestGreen}\textbf{0.03}} & -1.00 & {\textbf{-0.06}} & -0.75 & -0.77 \\
      \rowcolor{RowAlt}
      Omega
      & 0.58 & 1.05 & 0.76 & 0.68 & 0.84 & {\color{ForestGreen}\textbf{1.80}} & 0.62 & 0.97 & 0.76 & {\textbf{1.27}} &  & 0.08 & 0.51 & 0.21 & 0.27 & 0.22 & {\color{ForestGreen}\textbf{1.03}} & 0.17 & {\textbf{0.94}} & 0.32 & 0.41 \\
      MaxDD
      & -8.19 & -3.85 & -7.38 & -10.87 & -7.81 & {\color{ForestGreen}\textbf{-0.45}} & -8.73 & {\textbf{-3.29}} & -10.08 & -4.50 &  & -38.87 & -5.80 & -15.56 & -10.77 & -21.27 & {\color{ForestGreen}\textbf{-0.90}} & -18.58 & {\textbf{-2.36}} & -5.72 & -4.55 \\
      \rowcolor{RowAlt}
      Edge
      & 0.17 & 0.20 & {\textbf{0.46}} & 0.17 & 0.01 & 0.00 & 0.10 & 0.00 & {\color{ForestGreen}\textbf{1.14}} & 0.10 &  & 0.00 & 0.00 & 0.04 & 0.04 & {\textbf{0.38}} & 0.00 & 0.08 & {\color{ForestGreen}\textbf{0.63}} & 0.37 & 0.01 \\

    \addlinespace[0.5em]
    \midrule
    \addlinespace[0.5em]
    \multicolumn{22}{l}{\textcolor{dpd2}{\textbf{Panel B: Absolute Error}}} \\
    \addlinespace[0.3em]

      Return
      & 0.00 & {\textbf{0.04}} & -0.01 & -0.00 & {\color{ForestGreen}\textbf{0.05}} & 0.03 & -0.03 & 0.00 & 0.04 & 0.03 &  & -0.67 & -0.17 & -0.49 & -0.29 & -0.33 & {\color{ForestGreen}\textbf{0.02}} & -0.39 & {\textbf{-0.01}} & -0.14 & -0.14 \\
      \rowcolor{RowAlt}
      Sharpe
      & 0.01 & {\textbf{0.24}} & -0.03 & -0.02 & 0.23 & {\color{ForestGreen}\textbf{0.29}} & -0.14 & 0.01 & 0.12 & 0.17 &  & -0.78 & -0.41 & -0.88 & -0.75 & -0.67 & {\color{ForestGreen}\textbf{0.13}} & -0.97 & {\textbf{-0.05}} & -0.57 & -0.59 \\
      Sortino
      & 0.02 & {\textbf{0.41}} & -0.04 & -0.02 & 0.33 & {\color{ForestGreen}\textbf{0.49}} & -0.16 & 0.02 & 0.17 & 0.22 &  & -0.75 & -0.55 & -0.93 & -0.83 & -0.74 & {\color{ForestGreen}\textbf{0.20}} & -0.96 & {\textbf{-0.06}} & -0.66 & -0.75 \\
      \rowcolor{RowAlt}
      Omega
      & 1.02 & {\textbf{1.37}} & 0.96 & 0.97 & 1.36 & {\color{ForestGreen}\textbf{1.49}} & 0.80 & 1.02 & 1.17 & 1.29 &  & 0.15 & 0.58 & 0.32 & 0.37 & 0.43 & {\color{ForestGreen}\textbf{1.20}} & 0.30 & {\textbf{0.93}} & 0.49 & 0.48 \\
      MaxDD
      & -3.11 & {\color{ForestGreen}\textbf{-0.34}} & -0.89 & -4.38 & -0.95 & {\textbf{-0.42}} & -4.73 & -2.40 & -4.54 & -0.86 &  & -11.15 & -3.02 & -6.33 & -4.96 & -7.69 & {\color{ForestGreen}\textbf{-0.62}} & -7.89 & {\textbf{-1.54}} & -2.52 & -2.79 \\
      \rowcolor{RowAlt}
      Edge
      & 0.34 & 0.19 & {\textbf{0.57}} & 0.36 & 0.07 & 0.02 & 0.05 & 0.00 & {\color{ForestGreen}\textbf{1.84}} & 0.06 &  & 0.00 & 0.06 & 0.15 & 0.13 & {\color{ForestGreen}\textbf{0.92}} & 0.07 & 0.18 & 0.26 & {\textbf{0.44}} & 0.08 \\

    \addlinespace[0.5em]
    \midrule
    \addlinespace[0.5em]
    \multicolumn{22}{l}{\textcolor{dpd2}{\textbf{Panel C: Classical Forecast Accuracy}}} \\
    \addlinespace[0.3em]

      RMSE
      & 1.05 & 1.00 & 1.03 & 1.05 & 1.02 & {\color{ForestGreen}\textbf{0.98}} & 1.04 & 1.00 & 1.04 & {\textbf{0.98}} &  & 1.80 & 1.13 & 1.44 & 1.27 & 1.45 & {\color{ForestGreen}\textbf{1.00}} & 1.39 & {\textbf{1.01}} & 1.15 & 1.11 \\
      \rowcolor{RowAlt}
      MAE
      & 1.00 & {\textbf{0.96}} & 1.01 & 1.00 & {\color{ForestGreen}\textbf{0.95}} & 0.97 & 1.03 & 1.00 & 0.96 & 0.97 &  & 1.67 & 1.17 & 1.49 & 1.29 & 1.33 & {\color{ForestGreen}\textbf{0.98}} & 1.39 & {\textbf{1.01}} & 1.14 & 1.14 \\
      $\rho(1)$
      & 0.43 & 0.42 & 0.38 & 0.44 & {\textbf{0.36}} & 0.40 & 0.44 & 0.42 & 0.46 & {\color{ForestGreen}\textbf{0.28}} &  & {\color{ForestGreen}\textbf{0.06}} & 0.25 & 0.19 & 0.29 & 0.42 & 0.15 & 0.22 & {\textbf{0.09}} & 0.24 & 0.23 \\
      \rowcolor{RowAlt}
      DM $t$-stat
      & -1.19 & 0.09 & -0.54 & -0.79 & -0.37 & {\color{ForestGreen}\textbf{1.46}} & -0.80 & -0.07 & -0.59 & {\textbf{0.35}} &  & -1.15 & -1.06 & -1.94 & -2.95 & -2.06 & {\color{ForestGreen}\textbf{0.06}} & -2.22 & {\textbf{-0.10}} & -1.73 & -2.61 \\

    \addlinespace[0.3em]
    \bottomrule
  \end{tabular}%
  }

  \vspace{0.25em}
  \parbox{\linewidth}{\scriptsize
    \textit{Notes}: Panels A--B report risk-adjusted metrics; Panel C reports classical forecast accuracy metrics. $\rho(1)$ = first-order autocorrelation of errors.
    Best: \textcolor{ForestGreen}{\textbf{bold green}}; second-best: \textbf{bold}.
  }
  }

\end{table}
\end{landscape}





\end{document}